\documentclass[aps,pra,twocolumn,showpacs,floatfix,superscriptaddress]{revtex4-1}
\usepackage[utf8]{inputenc}
\usepackage[T1]{fontenc}
\usepackage[english]{babel}
\usepackage{float}
\usepackage{color}
\usepackage{graphicx}
\usepackage{indentfirst}
\usepackage{amsmath,amsfonts,amsthm,bm}
\begin{document}
\title{
Spin-spiral formalism based on the multiple scattering Green's function technique with applications to ultrathin magnetic films and multilayers
}
\author{E. Simon}
\email{esimon@phy.bme.hu}
\affiliation{Department of Theoretical Physics, Budapest University of Technology and Economic, Budafoki \'{u}t 8, H-1111 Budapest, Hungary}
\author{L. Szunyogh}
\affiliation{Department of Theoretical Physics, Budapest University of Technology and Economic, Budafoki \'{u}t 8, H-1111 Budapest, Hungary}
\affiliation{MTA-BME Condensed Matter Research Group, Budapest University of Technology and Economics, Budafoki \'{u}t 8, H-1111 Budapest, Hungary}

\begin{abstract}
Based on the Korringa-Kohn-Rostoker Green's function technique we present a computational scheme for calculating the electronic structure of layered systems with homogeneous spin-spiral magnetic state. From the self-consistent non-relativistic calculations the total energy of the spin-spiral states is determined as a function of the wave vector,  while a relativistic extension of the formalism in first order of the spin-orbit coupling gives an access to the effect of the Dzyaloshinskii-Moriya interactions. We demonstrate that the newly developed method properly describes the magnetic ground state of a Mn monolayer on W(001) and that of a Co monolayer on Pt(111). The obtained spin-spiral energies are mapped to a classical spin model, the parameters of which are compared to those calculated directly from the relativistic torque method.  In case of the Co/Pt(111) system we find that the isotropic interaction between the Co atoms is reduced and the Dzyaloshinskii-Moriya interaction is increased when capped by a Ru layer. In addition, we perform spin-spiral calculations on Ir/Fe/Co/Pt and Ir/Co/Fe/Pt multilayer systems and find a spin-spiral ground state with very long wavelength due to the frustrated isotropic couplings between the Fe atoms, whereas the Dzyaloshinskii-Moriya interaction strongly depends on the sequence of the Fe and Co layers.
\end{abstract}

\maketitle

\section{Introduction}
Applications of complex magnetic structures in modern information technology are for several decades in the focus of broad research interest \cite{Parkin,Fert, Duine2018}. %Beyond the experimental investigations, the theoretical description of the novel magnetic structures plays an increasingly important role. 
Advanced numerical simulatons based on first principles methods \cite{Bihlmayer} play an essential role for understanding the magnetic phenomena on a broad scale and also to design new devices. To this purpose classical spin models \cite{Nowak} are also used extensively where the parameters of the spin models can be derived from first principles calculations for collinear or non-collinear magnetic states. 

Spin-spiral states form a reach subset of non-collinear magnetic configurations that exist in nature, in particular, in thin magnetic films \cite{Bergmann_2014} and can be studied theoretically with analytical and computational tools. 
The first \textit{ab initio} approach for calculating the electronic structure in the presence of a spin-spiral magnetic state in terms of the generalized Bloch theorem was introduced by Sandratskii \cite{Sandratskii_PhysStatSol, Sandratskii_1991} and its implementation in the linearized muffin-tin orbital (LMTO) method was successfully used to calculate the energies of spin-spiral states in bulk Fe  \cite{Sandratskii_1986, Mryasov_1991}. Using the augmented spherical wave (ASW) technique it was possible to study the static non-uniform spin-susceptibiity of various bulk systems \cite{Sandratskii_AdvPhys}. 
%Fully unconstrained approach to non-collinear magnetism that treats the magnetization density as a continuous vector quantity can be also developed by various groups and used small atomic clusters \cite{PhysRevLett.76.4420, PhysRevLett.80.3622, PhysRevB.62.11556}. 
The calculation of spin-spiral states was implemented in the full-potential linearized augmented plane-wave (FLAPW) method as combined with a constrained local moment treatment and used to determine the electronic structure of  spin-spiral states in bcc Fe \cite{PhysRevB.69.024415}. 
In the spin-spiral calculations based on the LMTO,  ASW or FLAPW band structure methods an appropriate basis function set is used to solve the Kohn-Sham equations. % In the literature only one work can be found for the spin-spiral calculation in terms of multiple scattering theory, but  for non-relativistic case and this technique was applied for bulk system  \cite{PhysRevB.83.144401}.
On the contrary, a non-relativistic multiple scattering Green's function formalism  was developed for spin-spiral configurations in Ref.~\onlinecite{PhysRevB.83.144401} and applied to ordered and disordered solids with spiral magnetic order. 

For systems with broken space-inversion symmetry,  spin orbit coupling (SOC)  leads to the appearence of  Dzyaloshinskii-Moriya  interactions (DMI) \cite{Dzyaloshinsky,Moriya} that can stabilize  non-collinear chiral magnetic structures, such as spin spirals and magnetic skyrmions. %For this reason, the evaluation of the DMI in the system is a crucial task for the first principles calculations. 
The DMI can be determined from a collinear magnetic structure in terms of the relativistic torque method (RTM) \cite{rtm-Udvardi,rtm-Ebert,PhysRevB.82.100403,PhysRevB.99.104427} or from the spin-cluster expansion (SCE) technique based on the disordered paramagnetic state \cite{SCE-RDLM-1,SCE-RDLM-2}. The energy due to the DMI can also be obtained using a first order perturbation treatment of the spin-orbit coupling  on top of a non-relativistic spin-spiral calculation \cite{HEIDE20092678} or, at least for commensurate spin spirals, employing supercell calculations \cite{PhysRevLett.115.267210,PhysRevB.96.024450}.  %Expression of the DM coupling was reported based on Berry-phase approach \cite{Freimuth2014} and in terms of spin current \cite{PhysRevLett.116.247201}. Resolving DMI in reciprocal space, more insight into the dependence of the DM energy on the electronic structure was shown \cite{PhysRevB.96.024450}. 

The total energy as a function of the wave vector of the spin spirals can be mapped to Heisenberg model giving thus an accurate access to the isotropic coupling parameters in the system.  The main advantage of this procedure is  that the longitudinal fluctuations of the magnetic moments, including the induced moments, are taken into account \cite{Lezaic-PRB-2013}, while in case of non self-consistent approaches based on the magnetic force theorem \cite{Liechtenstein1987,PhysRevB.59.4699}, such as in the RTM os SCE, these longitudinal fluctuations are neglected. Morever, the self-consistent spin-spiral calculations include, in principle, all higher order magnetic exchange interactions. There are strong indications that these couplings can stabilize exotic complex magnetic states \cite{Heinze2011,PhysRevLett.120.207202}.

In this work we present the spin-spiral formalism within the multiple scattering Green's function technique for both the non-relativistic and the relativistic cases and, as implemented with the screened Korringa-Kohn-Rostoker (SKKR) method \cite{Szunyogh1,Zeller,Szunyogh2}, its applications to ultrathin magnetic films and multilayers. 
In Sec.~\ref{sec2} a detailed description of the non-relativistic theory is presented, together with a first-order perturbation technique to include spin-orbit induced effects, while we also give details of the spin model we use for thin magnetic films. The applications are presented in Sec.~\ref{sec3}.  We calculate the spin-spiral dispersion for a Mn monolayer on W(001) and find that the ground state is a right handed cycloidal spin-spiral state according to other theoretical and experimental results \cite{PhysRevLett.101.027201}. For a Co monolayer on Pt(111) we obtain that the ground state is ferromagnetic due to the large isotropic exchange coupling between the Co atoms and that the preferred rotational sense of the in-plane DM vector is left-handed,  in agreement with previous theoretical works \cite{Dupe2014,PhysRevB.94.214422,PhysRevB.99.214426} and with experiment \cite{Hrabec-PRB-2014}. We show that a Ru overlayer reduces the Co-Co exchange coupling and increases the in-plane DM interaction, thus the Ru/Co/Pt layer sequence can be an important component of the novel functional multilayer structures \cite {PhysRevMaterials.3.041401, Han2019}.  Room temperature skyrmions were observed in Ir/Fe/Co/Pt multilayers  \cite{Soumyanarayanan2017, Yagil_APL, Raju2019} and stable skyrmionic states in $4d$/Fe$_{2}$/$5d$ multilayers were predicted theoretically \cite{Dupe2016}. Motivated by these experimental and theoretical works we investigate Ir/Fe/Co/Pt and Ir/Co/Fe/Pt multilayer systems and highlight that the DMI energy strongly depends on the sequence of the Fe and Co layers. Finally in Sec.~\ref{sec:summary} we summarize our results and draw possible conclusions.

\section{Theoretical background \label{sec2}}
\subsection{Non-relativistic Green's function technique for spin-spiral states}

\begin{figure} [htb!]
\centering
\includegraphics[width=1.0\columnwidth]{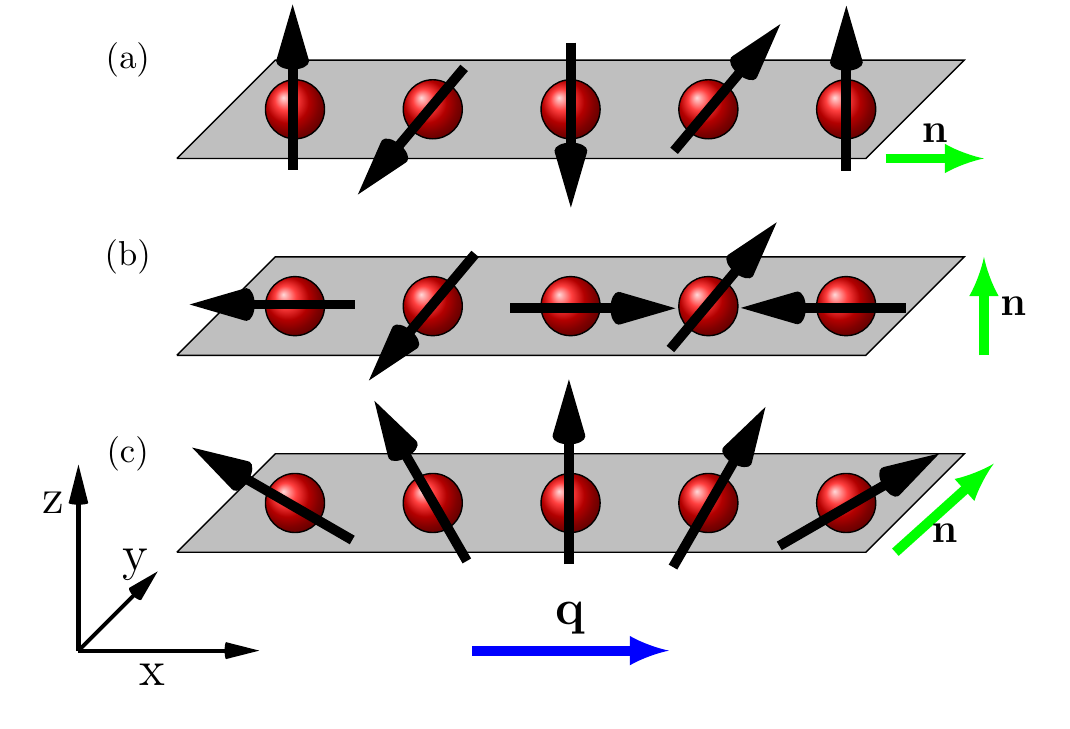}\\
\caption{Schematic representation of spin spirals with propagation vector $\mathbf{q}$ along the $x$ axis and with different rotational axes $\mathbf{n}$. In case of (a) $ \mathbf{q} \parallel \mathbf{n}$ a Bloch-type spin spiral, while in case of  $ \mathbf{q} \perp \mathbf{n}$ (b) an in-plane or (c) an out-of-plane N\'eel-type spin spiral are formed.}
\label{ss_draw}
\end{figure}
%The spins are rotated by a constant angle from atom to atom along a certain direction of the crystal in a magnetic structure is called spin-spiral. 

In this Section we present the non-relativistic multiple scattering formalism for homogeneous spin spirals along the lines of Ref.~\onlinecite{PhysRevB.83.144401}. 
Let us consider a lattice with translation vectors $\mathbf{T}_{n}$ and basis vectors $\mathbf{a}_\nu$ ($\nu=1,\dots,M$) defining the inequivalent sublattices of the lattice. For layered systems $\mathbf{T}_{n}$ are two-dimensional (2D) vectors, while $\mathbf{a}_\nu$ point to different layers. Note that for complex lattices the third  ($z$) component of  $\mathbf{a}_\nu$ should not necessarily be different for each of the layers.   
In a homogeneous spin spiral the spin-magnetic moments, 
\begin{equation}
\mathbf{M}_{\nu n}=M_{\nu}\,\mathbf{S}_{\nu n} \, .
\end{equation}
differ only in their orientations $\mathbf{S}_{\nu n}$ as
\begin{align}
\mathbf{S}_{\nu n} = {\cal{R}}(\mathbf{n}, \phi_n) \, \mathbf{S}_{\nu 0} \, ,
\label{snrot}
\end{align}
where ${\cal{R}}(\mathbf{n}, \phi_n) $ (later on denoted by ${\cal{R}}_n$) stands for a proper rotation around the axis defined by the unit vector $\mathbf{n}$ by an angle $\phi_n=\mathbf{qT}_{n}$, where $\mathbf{q}$ is the propagation vector of the spin spiral.
If at the reference site of sublattice $\nu$, corresponding to $\mathbf{T}_{0}=\mathbf{0}$, the direction of the spin  is  
\begin{align} \mathbf{S}_{\nu 0}=  \mathbf{n}\cos\theta_{\nu} \: + & \left[ \, \mathbf{e}_{t1} \cos\left(\varphi_{\nu}\right)  + \mathbf{e}_{t2}\sin\left( \varphi_{\nu}\right)  \right]  \sin\theta_{\nu} \, ,
\end{align}  
where $\mathbf{e}_{t1}\perp\mathbf{n}$ and $\mathbf{e}_{t2}=\mathbf{n}\times\mathbf{e}_{t1}$, 
  $\mathbf{S}_{\nu n}$ can be expressed as
\begin{align} \mathbf{S}_{\nu n}=  \mathbf{n}\cos\theta_{\nu} \: + & \left[ \; \: \mathbf{e}_{t1} \cos\left(  \mathbf{qT}_{n}+\varphi_{\nu}\right)  \right. \nonumber \\
 & \left. + \mathbf{e}_{t2}\sin\left(  \mathbf{qT}_{n}+\varphi_{\nu}\right)  \right]  \sin\theta_{\nu} \, .
\end{align}

For layered magnetic systems the propagation vector $\mathbf{q}$ lies in the plane of the layers, usually chosen the $(x,y)$ plane of the global frame of reference, and the rotational axis $\mathbf{n}$ is an arbitrary unit vector.
Various types of spin spirals are distinguished according to the relative direction of  $\mathbf{q}$ and $\mathbf{n}$.  Fig. \ref{ss_draw} illustrates a Bloch-type spin spiral with $\mathbf{q}\parallel \mathbf{n}$, as well as an in-plane and an out-of-plane N\'eel-type spin spiral with $\mathbf{q}\perp \mathbf{n}$. For these spin configurations we choose $\theta_\nu=\pi/2$, consequently, $\mathbf{S}_{\nu n} \perp \mathbf{n}$, therefore, they represent flat spin spirals.     

As the charge density and the magnitude of the magnetization densitiy are identical for each sites in a sublattice, within the local density approximation (LDA) of the density functional theory, the effective potential $V \left(  \mathbf{r}\right)$ and exchange field $\mathbf{B}\left(  \mathbf{r}\right)$ also preserves the translational-rotational symmetry,
\begin{equation}
V_{\nu n}\left(  \mathbf{r}\right)  =V_{\nu}\left(  \mathbf{r}\right) \, ,
\end{equation}%
\begin{equation}
\mathbf{B}_{\nu n}\left(  \mathbf{r}\right)  =\mathbf{S}_{\nu n}\,B_{\nu
}\left(  \mathbf{r}\right)  =\mathcal{R}_{n}\,\mathbf{S}_{\nu 0}\,B_{\nu
}\left(  \mathbf{r}\right)  \, .
\end{equation}
The Kohn-Sham equation in cell $n$ of sublattice $\nu$ can then can be written as:
\begin{align}
& \left(  -\Delta +V_{\nu }\left(  \mathbf{r} \right)+
\mu_{B} \, \left(  \mathcal{R}_{n}^{-1} \mbox{\boldmath $\sigma$}\right)  \cdot \mathbf{S}_{\nu 0} \, B_{\nu}\left(  \mathbf{r}\right)  
 \right)  \psi_{\nu n}\left(\mathbf{r} \right) \nonumber \\
& \ \qquad \qquad \qquad  =\varepsilon\,\psi_{\nu n}\left(  \mathbf{r}\right) ,
\end{align}
where we used atomic (Rydberg) units ($\hbar=1$, $2m=1$) and $\mbox{\boldmath $\sigma$}=(\sigma_x,\sigma_y,\sigma_z)$ denote the Pauli matrices. 
The transformation of the Pauli matrices can be expressed as%
\begin{equation}
\mathcal{R}_{n}^{-1}\bm{\sigma} = U_{n}^{+}\,\bm{\sigma\,}U_{n}, 
\end{equation}
where the unitary 2$\times$2 matrix $U_{n}$ is given by
\begin{equation}
U_{n}=\exp\left(  \frac{i}{2}  \mathbf{n} \bm{\sigma}\phi_{n}   \right)=    \exp\left(  \frac{i}{2}\left(  \mathbf{n} \bm{\sigma}%
\right)  \left( \mathbf{qT}_{n}\right)\right) .
\label{su2}
\end{equation}
Thus the Kohn-Sham equation for cell $\nu n$ takes the form,
\begin{align}
&\,\left(  -\Delta + V_{\nu}\left(  \mathbf{r}\right) 
+ \mu_{B}\,\mathbf{\sigma\cdot S}_{\nu 0}\,B_{\nu}\left(  \mathbf{r}\right) \right) U_n  \psi_{\nu n}\left(  \mathbf{r}\right) \nonumber \\
&  \ \qquad \qquad \qquad  = \varepsilon\,U_{n} \psi_{\nu n}\left(  \mathbf{r}\right)
\label{sol_sch}
\end{align}
which implies that if $\psi_{\nu n}\left(  \mathbf{r}\right)  $ is a solution of the Schr\"{o}dinger equation in cell $\nu n$, then $U_{n}\psi_{\nu n}\left(  \mathbf{r}\right)  $ is a solution for cell $\nu 0$.

Within the non-relativistic multiple scattering theory (MST) the $\left(  \ell \, m \, s\right)  =\left(  L \, s\right)$ angular momentum-spin representation is used with the free-space solutions $J^{Ls}(\varepsilon,\mathbf{r}) =j_{\ell}(\sqrt{\varepsilon} r) Y_{\ell m}(\hat r)  \phi_{s}$ and $N^{Ls}(\varepsilon,\mathbf{r}) =n_{\ell}(\sqrt{\varepsilon} r) Y_{\ell m}(\hat r)  \phi_{s}$, where $j_\ell(x)$ and $n_\ell(x)$ are the spherical Bessel- and Neumann functions, respectively, $Y_{\ell m}(\hat r)$ are spherical harmonics, while  $\phi_{s}$ denote spinor basis functions. Introducing the vector notation, $\underline{J}\left(  \varepsilon,\mathbf{r}\right) =\left\{  J^{Ls}\left(
\varepsilon,\mathbf{r}\right)  \right\}$ and $\mathcal{N}\left(  \varepsilon,\mathbf{r}\right)=\left\{  N^{Ls}\left(\varepsilon,\mathbf{r}\right)  \right\}$, the regular and irregular scattering solutions of the Kohn-Sham equation Eq. (\ref{sol_sch}) are normalized beyond the radius of the  atomic (muffin-tin) sphere as
\begin{equation}
\mathcal{Z}_{\nu n}\left(  \varepsilon,\mathbf{r}\right)  =\underline{J}\left(  \varepsilon,\mathbf{r}\right)
\underline{P}_{\nu n}\left(  \varepsilon\right)  +\sqrt{\varepsilon} \mathcal{N}\left(
\varepsilon,\mathbf{r}\right) \, ,
\label{regsol}
\end{equation}
and 
 \begin{equation}
\underline{J}_{\nu n}\left(  \varepsilon,\mathbf{r}\right)  =\underline{J}\left(  \varepsilon,\mathbf{r}\right) \, ,
\label{irregsol}
\end{equation}
respectively, where $\underline{P}_{\nu n}\left(  \varepsilon\right)$ is the inverse reactance matrix
in the $(L \,s)$ representation, which is connected to the single-site $t$-matrix $\underline{t}_{\nu n}\left(
\varepsilon\right) $ as 
\begin{equation}
\underline{P}_{\nu n}\left(  \varepsilon\right)  =\underline{t}_{\nu n}\left(
\varepsilon\right)  ^{-1}-i\sqrt{\varepsilon}\underline{I} \, ,
\end{equation}
with $\underline{I}$ denoting the unit matrix.
According to Eq.~(\ref{sol_sch}) $U_{n}\,Z^{L \, s}_{\nu n}\left(  \varepsilon,\mathbf{r}\right)$ is a solution of the Schr\"{o}dinger equation at site $\nu 0$, therefore it must be a linear combination of the functions $Z^{L \, s}_{\nu 0}\left(  \varepsilon,\mathbf{r}\right)$.  Taking into account the boundary condition Eq.~\eqref{regsol} this implies 
\begin{align}
U_{n}\, \mathcal {Z}_{\nu n}\left(  \varepsilon,\mathbf{r}\right) & =  \mathcal {Z}_{\nu 0}\left(  \varepsilon,\mathbf{r}\right) \underline{U}_{n} \, ,
\label{ztrans}
\end{align}
and similar for the functions $\mathcal {J}_{\nu n}\left(  \varepsilon,\mathbf{r}\right)$, together with the transformation for the inverse reactance matrices,
\begin{equation}
\underline{P}_{\nu n}\left(  \varepsilon\right)  =\underline{U}_{n}^{+}\,\underline{P}_{\nu 0}\left(  \varepsilon\right)  \,\underline{U}_{n} \, ,
\label{sol_trans}
\end{equation}
where we introduced the matrix $\underline{U}_{n}=\{ \delta_{LL'} U_{n}^{ss'} \}$. Consequently, the same transformation applies to the single-site $t$-matrices,
\begin{equation}
\underline{t}_{\nu n}\left(  \varepsilon\right)  =\underline{U}%
_{n}^{+}\,\underline{t}_{\nu 0}\left(  \varepsilon\right)  \,\underline{U}_{n}%
\,.
\label{t_trans}
\end{equation}

\ 

The following matrices in composite site-angular momentum-spin space, 
\begin{equation}
\underline{\underline{t}}\left(  \varepsilon\right) = \{ \delta_{\nu \mu} \delta_{nm} \underline{t}_{\nu n}\left(  \varepsilon\right) \}
\end{equation}
and
\begin{equation}
\underline{\underline{G}}_0\left(  \varepsilon\right) = \{ \underline{G}_{0,\nu n,\mu m}\left(  \varepsilon\right) \}
\label{g0}
\end{equation}
are used to calculate the matrix of the scattering-path operator (SPO),
\begin{equation}
\underline{\underline{\tau}}\left(  \varepsilon\right)   
=\left[ \underline{\underline{t}}\left(  \varepsilon\right)
^{-1}-\underline{\underline{G}}_{0}\left(  \varepsilon\right)
\right]  ^{-1} \, .
\label{eq_SPO}
\end{equation}
In Eq.~\eqref{g0} $\underline{G}_{0,\nu n,\mu m}\left(  \varepsilon\right)= \{ \delta_{ss'}  {G}^{LL'}_{0,\nu n,\mu m}\left(  \varepsilon\right)  \}$ denote the spin-independent free-space (bare) structure constants. 

Utilizing the transformation of the $t$-matrices Eq.~\eqref{t_trans}, the SPO matrix can be expressed as
\begin{align}
\underline{\tau}_{\nu n,\mu m}\left(  \varepsilon\right)   &  =\underline
{U}_{n}^{+}\,\underline{\widetilde{\tau}}_{\nu n,\mu m}\left(  \varepsilon
\right)  \,\underline{U}_{m},
\label{spo_trans}
\end{align}
where the $\underline{\widetilde{\tau}}_{\nu n,\mu m}\left(  \varepsilon\right)$ matrices can be calculated as
\begin{equation}
\underline{\underline{\widetilde{\tau}}}\left(  \varepsilon\right)   
=\left[ \underline{\underline{\widetilde{t}}}\left(  \varepsilon\right)
^{-1}-\underline{\underline{\widetilde{G}}}_{0}\left(  \varepsilon\right)
\right]  ^{-1} \, ,
\label{SPO-tilde}
\end{equation}
with $\underline{\underline{\widetilde{t}}}\left(  \varepsilon\right)$ comprising $t$-matrices that are identical within each sublattice
\begin{equation}
\underline{\underline{\widetilde{t}}}\left(  \varepsilon\right) = \{  \delta_{\nu \mu} \delta_{nm}
\underline{t}_{\nu 0}\left(  \varepsilon\right) \} \, ,
\label{tmat-tilde}
\end{equation}
and 
\begin{equation}
\underline{\underline{\widetilde{G}}}_{0}\left(  \varepsilon\right) =
\ \{  \underline{U}_{n}  \underline{G}_{0,\nu n,\mu m}\left(  \varepsilon\right) \underline{U}_{m}^{+}
 \} \, .
\label{g0-tilde}
\end{equation}

Exploiting the lattice Fourier transform of the free-space structure constants and Eq.~\eqref{su2},
the above quantity can be expressed  
\begin{align}
& \underline{U}_{n}  \underline{G}_{0,\nu n,\mu m}\left(  \varepsilon\right) \underline{U}_{m}^{+}   = \nonumber \\
& \qquad \qquad  \frac{1}{\Omega_{BZ}}\int\limits_{BZ}\,e^{i(\mathbf{T}_{m}-\mathbf{T}%
_{n})\mathbf{k}} \, \widetilde{\underline{G}}_{0,\nu\mu}(\mathbf{k},\varepsilon
)d^{d}k, 
\label{UGU}
\end{align}
where $d$=2 or 3 for two-dimensional or three-dimensional translational invariance, respectively, and $\Omega_{BZ}$ is the volume of the Brillouin zone. The matrix $\widetilde{\underline{G}}_{0,\nu\mu}(\mathbf{k},\varepsilon
)$ can be written in
spin space as
\begin{widetext}
\begin{equation}
\underline{\widetilde{G}}_{0,\nu\mu}(\mathbf{k},\varepsilon)=\frac{1}{2}\left(
\begin{array}
[c]{cc}%
\underline{G}_{0,\nu\mu}(\mathbf{k+}\frac{\mathbf{q}}%
{2},\varepsilon)(1-n_{3})+\underline{G}_{0,\nu\mu}(\mathbf{k-}\frac{\mathbf{q}}%
{2},\varepsilon)(1+n_{3})  & (n_{1}+in_{2})\left[  \underline
{G}_{0,\nu\mu}(\mathbf{k+}\frac{\mathbf{q}}{2},\varepsilon)-\underline
{G}_{0,\nu\mu}(\mathbf{k-}\frac{\mathbf{q}}{2},\varepsilon)\right] \\ \\
(n_{1}-in_{2})\left[  \underline{G}_{0,\nu\mu}(\mathbf{k+}%
\frac{\mathbf{q}}{2},\varepsilon)-\underline{G}_{0,\nu\mu}(\mathbf{k-}%
\frac{\mathbf{q}}{2},\varepsilon)\right]  &  \underline
{G}_{0,\nu\mu}(\mathbf{k+}\frac{\mathbf{q}}{2},\varepsilon)(1+n_{3})+\underline
{G}_{0,\nu\mu}(\mathbf{k-}\frac{\mathbf{q}}{2},\varepsilon)(1-n_{3})
\end{array}
\right)  \,,
\end{equation}
\end{widetext}
where in the above expression $\underline{G}_{0,\nu\mu}(\mathbf{k} , \varepsilon)$ stands for the lattice Fourier transform of the free-space structure constants that can be obtained via Ewald summation \cite{Ewald-1921,Ham-Segall-1961,Kambe1,Kambe2,Kambe3}. 
For the case of a rotational axis parallel to the $z$ axis  ($n_{1}=0$, $n_{2}=0$, $n_{3}=1$) $\underline{\widetilde{G}}_{0,\nu\mu}(\mathbf{k},\varepsilon)$ takes the diagonal form as reported in Ref. \onlinecite{PhysRevB.83.144401}.

Utilizing also the translational invariance of the $\widetilde{t}$-matrices in Eq.~\eqref{tmat-tilde}, the matrices $\underline{\widetilde{\tau}}_{\nu n,\mu m}\left(  \varepsilon\right)$ can be evaluted as
\begin{equation}
\underline{\widetilde{\tau}}_{\nu n,\mu m}\left(  \varepsilon\right) =
\frac{1}{\Omega_{BZ}}\int\limits_{BZ}\,e^{i(\mathbf{T}_{m}-\mathbf{T}%
_{n})\mathbf{k}} \, \widetilde{\underline{\tau}}_{\nu\mu}(\mathbf{k},\varepsilon)d^{d}k \, ,
\end{equation}
where $\widetilde{\underline{\tau}}_{\nu\mu}(\mathbf{k},\varepsilon)$ are the blocks of the inverse of the following matrix in sublattice-angular momentum-spin space,
\begin{equation}
\underline{\underline{\widetilde{M}}}(\mathbf{k},\varepsilon)=\left\{
\delta_{\nu \mu} \underline{{t}}_{\nu 0}(\varepsilon)^{-1}-
\underline{\widetilde{G}}_{0,\nu \mu}(\mathbf{k},\varepsilon)\right\}\, .
\end{equation}
The SPO-matrix can then be obtained from the transformation Eq.~\eqref{spo_trans}. 

The Green's function within the MST is given as
\begin{align}
& G\left(  \mathbf{r}+\mathbf{R}_{\nu n},\mathbf{r}'+\mathbf{R}_{\mu m};\varepsilon\right)  
 = \nonumber \\  
& \qquad \qquad \mathcal{Z}_{\nu n}\left(  \mathbf{r};\varepsilon\right)  \underline{\tau
}_{\nu n,\mu m}\left(  \varepsilon\right)  \mathcal{Z}_{\mu m}\left(
\mathbf{r}';\varepsilon\right)  ^{\times} \nonumber \\ 
&  \qquad \qquad - \delta_{\nu \mu} \delta_{nm} \:
\mathcal{Z}_{\nu n}\left(\mathbf{r}_<;\varepsilon\right)  \underline{J}_{\nu n}\left(  \mathbf{r}_>;\varepsilon\right)^{\times} \, ,%
\end{align}
with %the functions defined in Eqs.~\eqref{regsol} and \eqref{irregsol}, 
$\mathbf{r}_<$ and $\mathbf{r}_>$ denoting $\mathbf{r}$ 
or $\mathbf{r}'$ with the smaller and larger magnitude, respectively, while the superscript $\times$ stands for the functions when replacing $Y_L(\hat r) \phi_s$ by $Y_L(\hat r)^* \phi_s^+$ in Eqs.~\eqref{regsol} and \eqref{irregsol}. It is straightforward to show from Eqs. (\ref{ztrans}) and  (\ref{spo_trans}) that the site-diagonal Green's function transforms as
\begin{equation}
G\left(  \mathbf{r}+\mathbf{R}_{\nu n},\mathbf{r}'%
+\mathbf{R}_{\nu n};\varepsilon\right)  = U_{n} G\left(  \mathbf{r}%
+\mathbf{R}_{\nu 0},\mathbf{r}'+\mathbf{R}_{\nu 0};\varepsilon\right) U_{n}^{+} \, , %
\label{gftrans}
\end{equation}
which immediately implies that the charge and magnetization densities are the same in each atomic cell of a sublattice and the spin-magnetic moments rotate from site to site according to Eq.~\eqref{snrot}. This means that the homogeneous spin-spiral state treated in terms of generalized Bloch theorem is consistent with the non-relativistic density functional theory and a self-consistent calculation for a given wave-vector ${\mathbf q}$ can be performed on the cost of a calculation of a periodic collinear magnetic state.  It should be noted that if ${\mathbf S}_{\nu 0} \perp {\mathbf n}$ does not apply, i.e. in case of conical spin spirals, the spin configuration does not correspond to a stationary state, therefore, the self-consistent electronic structure should be determined by exerting appropriate transversal constraint to the local moments.   
   
%Using self-consistent procedure the total energy, E$_{\rm tot}$ of the system in the spin-spiral state is calculated as a function of the $\mathbf{q}$ vector from the two dimensional Brillouin zone.  
%
%
\subsection{Relativistic correction to the spin-spiral energy}
Within a relativistic theory the above formalism can not be used for calculating the electronic structure of a spin-spiral state.  Formally, one can see this by using the fact that, even neglecting orbital polarization effects, the operator $W_n =\exp\left(  \frac{i}{\hbar}\left(  \mathbf{n} \bm{J} \right)  \left( \mathbf{qT}_{n}\right)\right)$, $\bm{J}=\bm{L}+\bm{S}$ being the total angular momentum operator, must be used to describe the rotation of the spin-moments, which precludes to express the structure constants, $\widetilde{\underline{G}}_{0,\nu n,\mu m}(\varepsilon)$,  via lattice Fourier transformation as in Eq.~\eqref{UGU}. In more feasible terms, the magnetic anisotropy induced by the spin-orbit coupling (SOC) differentiates between different directions of the magnetic moments in the spin spiral, thus alters the electronic states from site to site. It is well-known also from spin-model simulations that in the presence of magnetic anisotropy an inhomogeneous spin spiral is formed \cite{Rocio-2015}.  Consequently, when including spin-orbit coupling, supercell calculations as based on constrained local moments are needed to treat homogeneous spin-spiral states self-consistently \cite{PhysRevLett.115.267210}. 

It is appropriate to treat the SOC as perturbation in spin-spiral calculations \cite{HEIDE20092678}. Here we present a numerical scheme based on the magnetic force theorem to include the effect of SOC in first order. 
Instead of calculating the total energy of the spin spiral, we will consider the zero-temperature grand potential of the electronic system %due to SOC \cite{PhysRevB.59.4699}. For noninteracting electrons, the grand potential can be expressed at zero temperature as
\begin{equation}
\Omega = \int\displaylimits_{-\infty}^{E_{F}} d\varepsilon(\varepsilon - E_{F})n(\varepsilon)=- \int\displaylimits_{-\infty}^{E_{F}}d\varepsilon N(\varepsilon) \, ,
\end{equation}
where $E_{F}$ denotes the Fermi energy, $n(\varepsilon)$ is the density of states and $N(\varepsilon)$ is the integrated density of states.
Employing Lloyd's formula \cite{Lloyd_1967}, the grand potential can be expressed within MST as 
\begin{equation}
\Omega = -\frac{1}{\pi}\operatorname{Im}\int\displaylimits_{-\infty}^{E_{F}}d\varepsilon\mathrm{Tr\,}\mathrm{ln\,}\underline{\underline{\tau}}(\varepsilon) \, .
\label{lloyd}
\end{equation}
Introducing %the matrix $\Delta\underline{\underline{m}}\left(  \varepsilon\right)$ being 
the difference between the inverse $t$-matrices in the presence of SOC, $\underline{\underline{t}}^\prime\left(  \varepsilon \right)$, and in the absence of SOC, $\underline{\underline{t}}\left(\varepsilon\right)$,
\begin{equation}
\Delta\underline{\underline{m}}\left(  \varepsilon\right)
=\underline{\underline{t}}^\prime\left(  \varepsilon\right)
^{-1}-\underline{\underline{t}} \left(\varepsilon\right)  ^{-1},
\end{equation}%
 the change of the grand potential to first order in $\Delta\underline{\underline{m}}\left(  \varepsilon\right)$ can be expressed as, see Ref. \cite{rtm-Udvardi},
\begin{align}
\Delta \Omega \simeq & \frac{1}{\pi}\operatorname{Im} \int\displaylimits_{-\infty}^{E_{F}} d\varepsilon \, \mathrm{Tr}\left(  \Delta
\underline{\underline{m}}\left(  \varepsilon\right)
\underline{\underline{\tau}}\left(  \varepsilon\right)  \right) \\
=&\frac{1}{\pi} \sum_{\nu n} \operatorname{Im}  \int\displaylimits_{-\infty}^{E_{F}} d\varepsilon \, \mathrm{Tr}\left(  \Delta
\underline{m}_{\nu n}\left(  \varepsilon\right)
\underline{\tau}_{\nu n, \nu n} \left(  \varepsilon\right)  \right)
\;,
\label{DOmega}
\end{align}
where $\underline{\tau}_{\nu n, \nu n} \left(  \varepsilon\right)$ are the SPO matrices related to the spin spiral within the non-relativistic theory. 

In the presence of SOC, we write the Hamiltonian in cell $\nu n$ as
\begin{equation}
H'_{\nu n}=H_{\nu n}+H^{\rm SOC}_{\nu} \, ,
\end{equation}%
where $H_{\nu n}$ is the non-relativistic Kohn-Sham Hamiltonian and, for spherical potentials, the Hamiltonian of the spin-orbit coupling is given by
\begin{equation}
H^{\rm SOC}_{\nu}=\xi_\nu (r) \,  \mathbf{LS} \, ,
\end{equation}%
with 
\begin{equation}
\xi_\nu (r)  =\frac{1}{2m^{2}c^{2}}\frac{1}{r}\frac{dV_{\nu}\left(
r\right)  }{dr}  \, ,
\end{equation}
where $\mathbf{L}$ and $\mathbf{S}$ denote the operators of the electron's angular momentum and spin, respectively.
In order to evaluate the change of the $t$-matrix due to SOC we use the regular solutions,
\begin{equation}
\mathcal{R}_{\nu n} \left(  \mathbf{r},\varepsilon\right) =
\mathcal{Z}_{\nu n} \left(  \mathbf{r},\varepsilon\right)  \underline{t}_{\nu n}\left(  \varepsilon\right) \, ,
\label{R-Zt}
\end{equation}
normalized beyond the radius of the atomic sphere $S_\nu$ as
\begin{equation}
\mathcal{R}_{\nu n} \left(  \mathbf{r},\varepsilon\right)=\underline{J}\left(  \mathbf{r},\varepsilon\right)  -i \sqrt{\varepsilon} \mathcal{H}\left(  \mathbf{r},\varepsilon\right) \underline{t}_{\nu n}\left(  \varepsilon\right) \, ,
\label{Rnorm}
\end{equation}
where $\mathcal{H}\left(  \mathbf{r},\varepsilon\right)$ are the Hankel-type solutions of the free-space Schr\"{o}dinger equation.  
The first-order (Born) approximation to the Lippmann-Schwinger equation for $\mathcal{R}^\prime_{\nu n} \left(  \mathbf{r},\varepsilon\right)$ then reads 
\begin{align}
& \mathcal{R}^\prime_{\nu n} \left(  \mathbf{r},\varepsilon\right) =\mathcal{R}_{\nu n}\left(  \mathbf{r},\varepsilon\right) 
\nonumber \\
& \quad+ \int\displaylimits_{r^{\prime}<S_\nu}  d^3r^{\prime} \, G_{\nu n}\left(  \mathbf{r},\mathbf{r}^{\prime},\varepsilon\right)
H^{\rm SOC}_{\nu} \left(  \mathbf{r}^{\prime}\right)  \mathcal{R}_{\nu n}\left(\mathbf{r}^{\prime},\varepsilon\right) \, ,
\label{Born}
\end{align}
where $G_{\nu n}\left(  \mathbf{r},\mathbf{r}^{\prime},\varepsilon\right)$ is the single-site Green's function, which for $r > S_\nu$
and $r'<S_\nu$ can be expressed as
\begin{equation}
G_{\nu n}\left(  \mathbf{r},\mathbf{r}^{\prime},\varepsilon\right) = -i \sqrt{\varepsilon} \mathcal{H}\left(  \mathbf{r},\varepsilon\right) 
\mathcal{R}_{\nu n}\left(\mathbf{r}^{\prime},\varepsilon\right)^\times \, .
\label{ssGF}
\end{equation}

 Inserting Eq.~\eqref{ssGF} into Eq.~\eqref{Born} we obtain a form like Eq.~\eqref{Rnorm} for $\mathcal{R}^\prime_{\nu n} \left(  \mathbf{r},\varepsilon\right)$ from which the change of the $t$-matrix can be read off,
\begin{align}
\Delta \underline{t}_{\nu n}\left(  \varepsilon\right) = \int\displaylimits%
_{r<S_\nu}  d^3r \, \mathcal{R}_{\nu n}\left(  \mathbf{r},\varepsilon\right)
^{\times}H^{\rm SOC}_{\nu} \left(  \mathbf{r}\right)  \mathcal{R}_{\nu n}\left(
\mathbf{r},\varepsilon\right)  \, .
\end{align}
Using the relationship Eq.~\eqref{R-Zt}, $\Delta \underline{m}_{\nu n}\left(  \varepsilon\right)$
can finally be expressed as
\begin{equation}
\Delta \underline{m}_{\nu n}\left(  \varepsilon\right) = - \int\displaylimits%
_{r<S_\nu}  d^3r \, \mathcal{Z}_{\nu n}\left(  \mathbf{r},\varepsilon\right)
^{\times}H^{\rm SOC}_{\nu} \left(  \mathbf{r}\right)  \mathcal{Z}_{\nu n}\left(
\mathbf{r},\varepsilon\right)  \, .
\label{deltam}
\end{equation}

We have to emphasize that the regular solutions $\mathcal{Z}_{\nu n} (\mathbf{r},\varepsilon)$ and the SPO matrices $\underline{\tau}_{\nu n, \nu n} (\varepsilon)$ entering Eqs.~\eqref{deltam} and \eqref{DOmega}, respectively, refer to the non-relativistic case, therefore, the transformations \eqref{ztrans} and \eqref{spo_trans} apply. From these it follows that
\begin{align}
& \mathrm{Tr}\left(  \Delta \underline{m}_{\nu n}\left(  \varepsilon\right)
\underline{\tau}_{\nu n, \nu n} \left(  \varepsilon\right)  \right) = \nonumber \\
& \qquad \qquad \mathrm{Tr}\left(  \Delta \underline{m}_{\nu 0}\left(  \varepsilon ; \mathbf{S}_{\nu n}\right)
\underline{\tau}_{\nu 0, \nu 0} \left(  \varepsilon ; \mathbf{S}_{\nu n} \right)  \right) \, ,
\label{Trn0-1}
\end{align}
where on the right-hand side we explicitely marked that at the reference site $\nu 0$ the orientation of the magnetization is changed from $\mathbf{S}_{\nu 0}$ to $\mathbf{S}_{\nu n}$. 
Here we have to emphasize that due to the first-order perturbation treatment of the SOC there is no magnetic anisotropy included in the energy $\Delta \Omega$. Consequently, such a change of the spin vector at the reference site shouldn't affect the contribution associated with site $\nu 0$. We tested numerically that, within a relative accuracy of about $10^{-5}$, the different sites in a sublattice add the same amount to the change of the energy (grand potential) due to the SOC, thus, it is sufficient to evaluate only the term for $\nu 0$ in Eq.~\eqref{DOmega} to obtain the  energy correction per site. 

We also note that it is possible to give an explicit expression for the SOC-induced energy per site, $\Delta \overline{\Omega}$. This is based on the reformulation of Eq.~\eqref{Trn0-1}  
\begin{equation}
\mathrm{Tr}\left(  \Delta \underline{m}_{\nu n}\left(  \varepsilon\right)
\underline{\tau}_{\nu n, \nu n} \left(  \varepsilon\right)  \right) =  \mathrm{Tr}\left(  \Delta \underline{\widehat m}_{\nu n}\left(  \varepsilon \right)
\underline{\tau}_{\nu 0, \nu 0} \left(  \varepsilon \right)  \right) \, ,
\label{Trn0-2}
\end{equation}
where $\Delta \underline{\widehat m}_{\nu n}\left(  \varepsilon \right)$ is defined by taking 
\begin{equation}
U_n^+ H^{\rm SOC}_{\nu} \left(  \mathbf{r}\right)  U_n = \xi_\nu (r) \,  \mathbf{L} \left( \mathcal{R}_n^{-1} \mathbf{S} \right)
\end{equation} 
instead of $H^{\rm SOC}_{\nu} \left(  \mathbf{r}\right) $ in Eq.~\eqref{deltam} with the scattering solutions $\mathcal{Z}_{\nu 0}\left(  \mathbf{r},\varepsilon\right)$. Averaging $\mathcal{R}_n^{-1} \mathbf{S}$ over $n$ for a wavelength of the spin spiral yields $\mathbf{n} \left(\mathbf{n}\mathbf{S}\right)$, where $\mathbf{n}$ refers to the axis of the rotation. Thus we arrive at
\begin{align}
\Delta {\overline \Omega} = \frac{1}{\pi} \sum_{\nu} \operatorname{Im}  \int\displaylimits_{-\infty}^{E_{F}} d\varepsilon \, \mathrm{Tr}\left(  \Delta \underline{\overline m}_{\nu}\left(  \varepsilon\right)
\underline{\tau}_{\nu 0, \nu 0} \left(  \varepsilon\right)  \right) \;,
\label{DOmave}
\end{align}
with
\begin{align}
& \Delta \underline{\overline m}_{\nu}\left(  \varepsilon\right) = \nonumber \\ 
&- \int\displaylimits%
_{r<S_\nu}  d^3r \, \mathcal{Z}_{\nu 0}\left(  \mathbf{r},\varepsilon\right)^{\times} 
\xi_\nu(r) \left(\mathbf{n}\mathbf{L}\right) \left(\mathbf{n}\mathbf{S}\right)
  \mathcal{Z}_{\nu 0}\left(\mathbf{r},\varepsilon\right)  \, .
\label{dmave}
\end{align}
Despite of the above closed expression for $\Delta {\overline \Omega}$, we used the $\nu 0$ term in Eq.~\eqref{DOmega} to calculate the DM energy per site of the spin spiral, since the evaluation of the matrix in Eq.~\eqref{deltam} is fairly simple when working in the relativistic $(j,\ell,m_j)$ representation. 

Because of the missing magnetic anisotropy $\Delta \Omega$ vanishes at $\mathbf{q}=0$ and it changes sign when reversing the wavevector, $\mathbf{q} \rightarrow -\mathbf{q}$, or, equivalently, the sense of rotation, $\mathbf{n} \rightarrow -\mathbf{n}$. $\Delta \Omega$ can therefore be identified with the DM energy of the spin spiral, $E_{\rm DM}$, which we add to the self-consistent non-relativistic total energy associated with the energy of isotropic spin-spin interactions. In addition, the expression Eq.~\eqref{dmave} should be correlated with the fact that only the components of the DM vectors being parallel to ${\mathbf n}$ contribute to the DM energy of a spin spiral.

\subsection{Spin-model parameters}
For comparison with the results of the spin-spiral calculations as outlined above, we will use the extended classical Heisenberg model,
\begin{equation}
\mathcal{H}=-\frac{1}{2}\sum_{ij}\mathbf{S}_{i}\underline{J}_{ij}\mathbf{S}_{j} - \sum_{i}\mathbf{S}_{i}\underline{K}_i\mathbf{S}_{i},
\label{hamiltonian}
\end{equation}
where $\underline{J}_{ij}$ is the exchange coupling tensor \cite{rtm-Udvardi}, and $\underline{K}_i$ is the on-site anisotropy matrix. The exchange coupling tensor can be decomposed into an isotropic, an antisymmetric and a traceless symmetric parts,
\begin{equation}
\underline{J}_{ij}=J_{ij}\underline{I}+\frac{1}{2}\left(\underline{J}_{ij}-\underline{J}^{T}_{ij}\right)+[\frac{1}{2} (\underline{J}_{ij}+\underline{J}^{T}_{ij})-J_{ij}\underline{I}],
\end{equation}
where $\underline{I}$ stands for the $3\times3$ unit matrix.
The isotropic part $J_{ij}=\frac{1}{3}\textrm{Tr}\underline{J}_{ij}$ represents the Heisenberg couplings between the magnetic moments. According to the sign convention of Eq. (\ref{hamiltonian}), $J_{ij} > 0$ and  $J_{ij} < 0$ indicate ferromagnetic (FM) and antiferromagnetic (AFM) couplings, respectively. The antisymmetric part of the exchange tensor can be identified with the DM vector $\mathbf{D}_{ij}$ as follows
\begin{equation}
\mathbf{S}_{i}\frac{1}{2}\left(\underline{J}_{ij}-\underline{J}^{T}_{ij}\right)\mathbf{S}_{j}=\mathbf{D}_{ij}\left(\mathbf{S}_{i}\times\mathbf{S}_{j}\right) \, .\label{eqn3}
\end{equation}
 The traceless symmetric part of the exchange tensor is related to the two-site magnetic anisotropy, while the second term on the right-hand side of Eq.~\eqref{hamiltonian} to the one-site magnetic anisotropy. In all cases considered in this work the symmetry of the system implied uniaxial on-site anisotropy $-\sum_i K_i S_{i,z}^2$.

The spin-model parameters in Eq.~(\ref{hamiltonian}) were determined by using the relativistic torque method as outlined in Ref.~\onlinecite{rtm-Udvardi}. %based on calculating the energy costs of infinitesimal rotations around 
Note that, in order to obtain all the matrixelements of the $3 \times 3$ exchange coupling matrices, ferromagnetic reference states oriented along different crystallographic directions should be used \cite{rtm-Udvardi}. In order to produce coupling matrices that respect the symmetry of the lattice, for these orientations we considered the out-of plane ($z$) direction and, in case of $C_{4v}$ and $C_{3v}$ point-group symmetry, two and three independent in-plane directions, respectively.

In order to facilitate a comparison between the spin-spiral calculations and the spin model, Eq.~\eqref{hamiltonian}, in case of fcc(111) and hcp(0001) surfaces we also determined the effective interaction parameters from the atomic interaction parameters derived for a large number of neighbors using RTM. According to Ref. \cite{PhysRevB.97.134405}, the effective spin-model parameters for $C_{3v}$ point group symmetry are defined as 
\begin{eqnarray}
J_{\rm eff}&=& \frac{1}{4} \sum_j J_{ij} (R^y_{ij})^2,\label{effpars1}
\\
D_{\rm eff}&=& \sum_j D^x_{ij} R^y_{ij} \, ,\label{effpars2}
\end{eqnarray}
known also as the spin stiffness \cite{Schweflinghaus-PRB2016} and spiralization \cite{Freimuth2014}, respectively.
The relationship between the micromagnetic and the effective parameters is given by
\begin{equation}
\mathcal{J} = \frac{1}{V_a} J_{\rm eff} \, , \; \mathcal{D}=\frac{1}{V_a} D_{\rm eff}   \;.
\label{micromagnpars}
\end{equation}
where $V_{a}=\frac{\sqrt{3}}{2}a_{\rm 2D}^2t$  is the atomic volume, $a_{\rm 2D}$ and $t$ being the in-plane lattice constant and the film thickness, respectively. Beyond the effective and micromagnetic parameters we can also define effective nearest neighbor interactions, $J$ and $D$, which are related to the effective parameters as:
\begin{equation}
J_{\rm eff}=  \frac{3}{4} a_{\rm 2D}^2 J \, ,  \quad  D_{\rm eff}= \frac{3}{2} a_{\rm 2D} D \, .
\label{effNNpars}
\end{equation}
In the present case we use the same sign convention as in Ref. \cite{PhysRevB.97.134405}, namely, $D > 0$ corresponds to the left-handed (counterclockwise) rotational direction, while $D < 0$ to the right-handed (clockwise) rotational direction. 
The effective spin-model parameters can obviously determined from the long-wavelength (small-wavenumber) limit of the spin-wave spectrum:  $J_{\rm eff}$ is related to the curvature of the non-relativistic energy dispersion, while $D_{\rm eff}$ to the slope of the SO induced contribution at $q=0$. 
From the calculated self-consistent total energy of the spin spirals  we also determined the isotropic exchange parameters for several neighbors using least-squares fit, where the isotropic couplings from the RTM served as initial parameters for the fitting procedure.  

In our applications we considered out-of-plane cycloidal spin spirals implying that all the spin vectors are in the ($\mathbf{q},\mathbf{z}$) plane. It is well-known that in this case the uniaxial anisotropy gives a contribution of $-K_i/2$ to the energy per site of a homogeneous spin spiral, while it adds  $-K_i$ to the energy of the ferromagnetic state along the $z$ direction with respect to an in-plane direction of the magnetization. We then approximated the missing magnetic anisotropy part to the energy of the spin spirals by adding half of the magnetoscrystalline anisotropy energy (MAE) calculated for the ferromagnetic system. We derived the MAE in the spirit of the  magnetic force theorem as a difference of the grand potential based on the band energy between the $x$ and $z$ direction of the magnetization, MAE = $\Omega_{x}- \Omega_{y}$ \cite{PhysRevB.59.4699, PhysRevB.41.11919, PhysRevB.51.9552}.  

In all calculations we used the atomic sphere approximation (ASA) for the effective potential with an angular momentum cut-off $\ell_{max}=2$ and the local spin-density approximation as parametrized by Vosko \textit{et al.}  \cite{vosko}. The energy integrals were performed along a semicircle contour in the upper complex energy semiplane. In case of the self-consistent calculations we used  $3300-3600$ $k$-points in the full Brillouin-zone, while for the calculation of the DM energy more than $12000$ $k$-points were necessary to achieve a reliable accuracy. 

\section{Results \label{sec3}}
\subsection{Mn monolayer on W(001)}
First we investigated the magnetic ground state of a Mn monolayer on W(001) in terms of spin-spiral calculations 
as implemented within the SKKR code for layered system. 
The model system consisted of four W layers, one Mn monolayer and three layers of empty spheres between a semi infinite W substrate and a semi infinite vacuum region. For the W and Mn layers epitaxial growth was assumed on a bcc(001) surface  with the in-plane lattice constant of W(001), $a_{2\rm D}=3.165$ \AA . The interlayer distance between the Mn layer and the topmost W layer was optimized by  VASP calculations \cite{Kresse199615, Kresse-PRB, Hafner}. Relative to the interlayer distance in bulk W we found an inward relaxation of 12.6 \%  for the Mn monolayer. 

\begin{figure} [htb!]
\centering
\includegraphics[width=1.0\columnwidth]{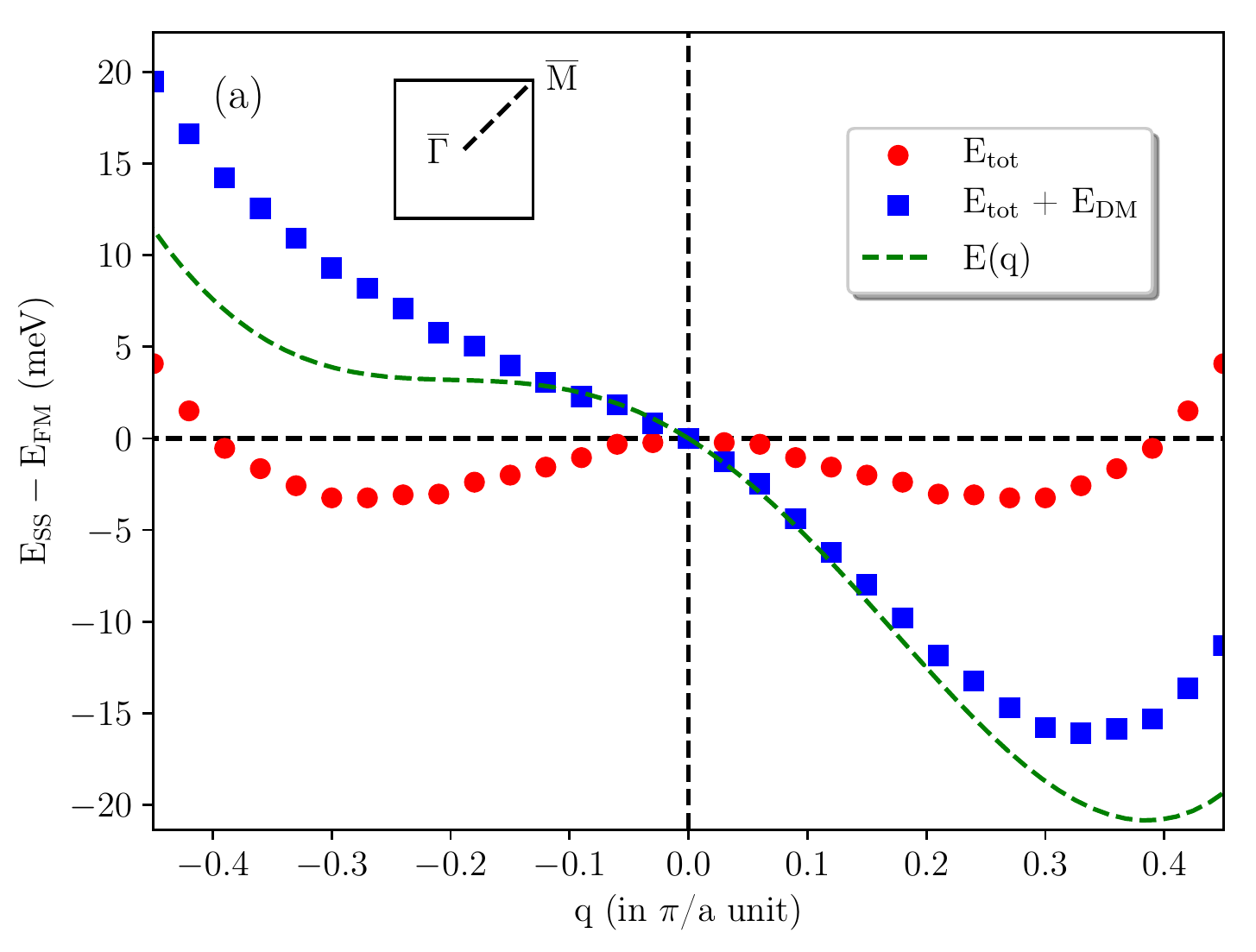}\\
\includegraphics[width=1.0\columnwidth]{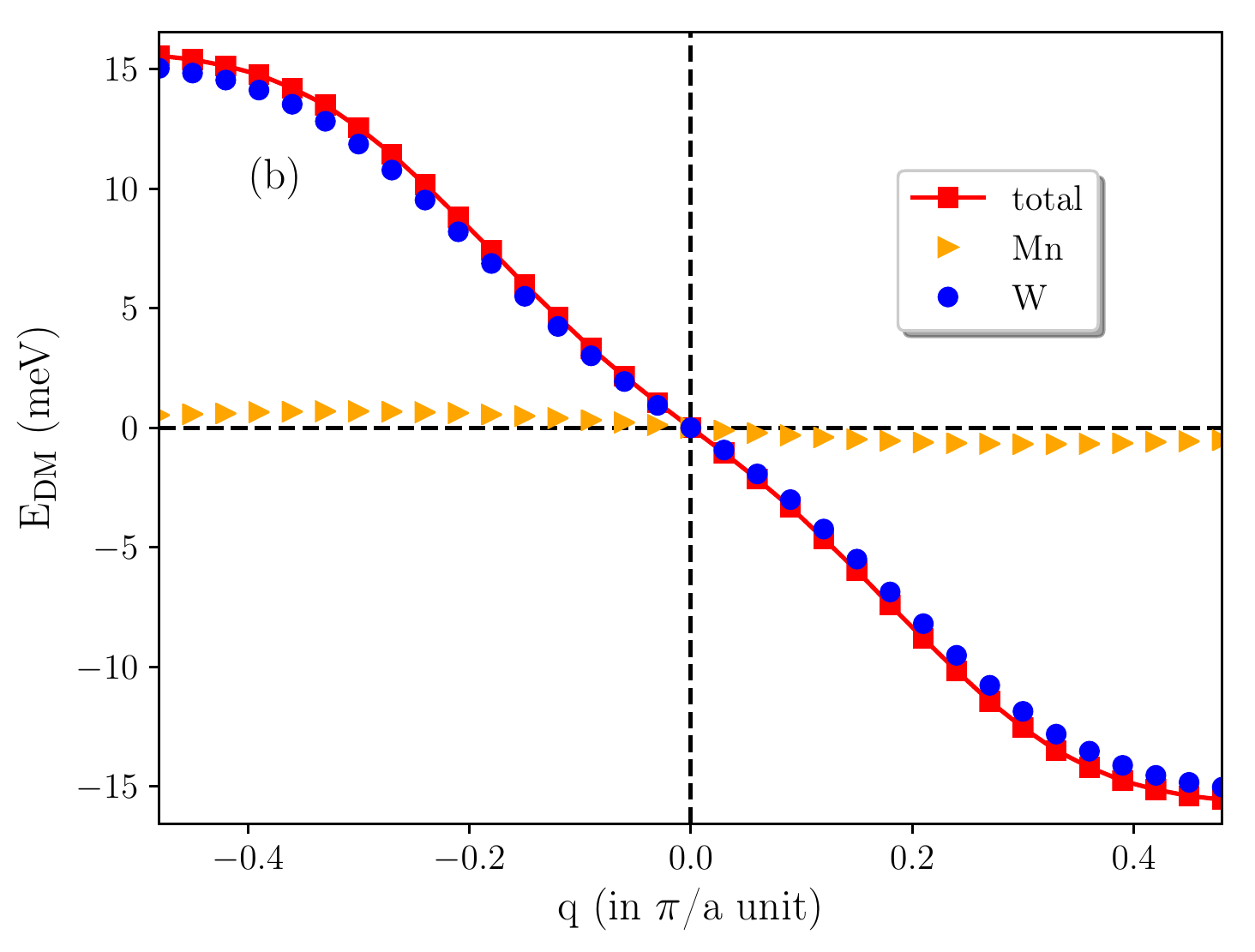}
\caption{(a) Calculated energies,  E$_{\rm SS}$ of out-of-plane homogeneous cycloidal spin spirals in a Mn monolayer on W(001) relative to the ferromagnetic state,  E$_{FM}$,  along the $\overline{\Gamma \rm M}$ direction in the Brillouin zone (see inset). The sum of the non-relativistic spin-spiral total energy, E$_{\rm tot}$ and the Dzyaloshinskii-Moriya energy,  E$_{\rm DM}$ implies a right-handed cycloidal spin-spiral as the magnetic ground state of the system. The dashed line represents the spin-spiral energy  E($q$) determined from the spin model in Eq. (\ref{hamiltonian}) using just the isotropic and DM coupling parameters calculated via the RTM. (b) The calculated DM energy of the spin spirals, see Eq.~\ref{DOmega}, with its decomposition into Mn and W contributions.}
\label{etot-mnw001}
\end{figure}

The total energy of homogeneous flat spin spirals, E$_{\rm tot}$ propagating along the (110) direction is shown in Fig.~\ref{etot-mnw001} (a) as a function of the wavenumber $q$. The magnitude of the Mn magnetic moment remained nearly constant with a value of 3.15 $\mu_{\rm B}$, while the induced moments of the W atoms changed in magnitude as a function of the wavenumber of the spin spiral. Note that at the $\overline{\rm M}$ point of the Brillouin zone, i.e. for a row-wise AFM configuration of the Mn moments, the induced moments of W vanishes. Apparently, as the state with lowest energy we obtained a spin spiral with a wavenumber of $q=0.3$ \AA$^{-1}$, thus, with a wavelength of $\lambda=2.1$~nm.  This is in relatively good agreement with the corresponding wavelength of $\lambda=3.1$~nm reported in Ref.~\onlinecite{PhysRevLett.101.027201} for the same system in terms of FLAPW spin-spiral calculations.  The energy gain of this spin spiral is about 3 meV/Mn atom with respect to the ferromagnetic state.

We fitted the calculated total energies of the spin spirals to an isotropic Heisenberg model containing the first five nearest neighbor (NN) interactions. We found that the fitted curve matched the calculated points very accurately along the whole $\overline {\Gamma \rm M}$ line. The fitted interactions are shown in Table \ref{mn-jiso-table}, together with those calculated by using the RTM. It should be mentioned that along this propagation direction of the spin spiral, the second and third NN couplings ($J_2$ and $J_3$) can not be obtained independently from a fitting up to the fifth NN interactions, since only $J_2+2J_3$ can uniquely be determined. We therefore present this value in Table \ref{mn-jiso-table}. 
Obviously, the dominant coupling is the strong ferromagnetic nearest-neighbor interaction, while the farther interactions are antiferromagnetic and considerably  smaller in size than the NN interaction. The interactions obtained from the two methods compare well to each other, except for the fifth NN coupling which is ferromagnetic for the RTM and antiferromagnetic in case of the spin-spiral fit. The isotropic interactions fitted to spin-spiral energies calculated by the FLAPW method \cite{PhysRevLett.116.177202} are also listed in Table~\ref{mn-jiso-table}. Note that due to the definition of the spin Hamiltonian in Eq. (\ref{hamiltonian}), the parameters given in Ref.~\onlinecite{PhysRevLett.116.177202} were multiplied by a factor of two. While there is an overall good agreement between the parameters obtained from the two spin-spiral calculations, $J_1$ and $J_2+2J_3$ from the FLAPW method are clearly smaller in magnitude than our calculated values. This can partly be attributed to the considerably smaller inward relaxation of the Mn monolayer used in the FLAPW calculations (4.7$\,$\%) \cite{PhysRevB.72.024452} as compared to our calculations (12.6$\,$\%). 
\begin{table}[t!]
\caption{Isotropic  exchange couplings for the first five nearest neighbors in a Mn monolayer on W(001) obtained by fitting of the spin-spiral energies to a Heisenberg model and from the relativistic torque method. Note that only $J_{2} +2  J_{3}$ could be determined from the fitting procedure. For comparison, the corresponding parameters reported in  Ref. \onlinecite{PhysRevLett.101.027201} are also shown as multiplied by two according to the spin model Eq. (\ref{hamiltonian}).}
\label{mn-jiso-table}
\centering
\begin{ruledtabular}
\begin{tabular}{l c c c c }
  method           &  $J_{1}$ & $J_{2} +2  J_{3}$    &  $J_{4}$       &      $J_{5}$         \\

\hline
   spin spiral        &      59.62        &      -29.27         &  -1.54  &       -0.55                            \\
    RTM              &        51.38       &       -26.23         &  -2.87   &       0. 81                           \\
   spin spiral \cite{PhysRevLett.116.177202} & 39.4  & -11.0 & -1.0 & -0.30                          \\
\end{tabular}
\end{ruledtabular}
\end{table}

The DM energy, E$_{\rm DM}$ calculated from Eq.~\eqref{DOmega} adds an antisymmetric term to the spin-spiral dispersion as shown in Fig. \ref{etot-mnw001}(b). As a result we find the energy minimum for a right-handed cycloidal spin-spiral with a period of $\lambda=1.92$ nm  which agrees well both with the experimental value of  $\lambda=2.1$ nm and with the value of $\lambda=2.3$ nm obtained from FLAPW spin-spiral calculations including SOC \cite{PhysRevLett.101.027201}. Reassuringly, the W atoms exhibiting large SOC and considerable spin-polarization have an overwhelming contribution to the DM energy, while the contribution of the Mn atoms is negligible, see Fig. \ref{etot-mnw001}(b).
For small $q$ values, the DM energy of the spin spirals is proportional with $q$ and from the slope of the curve an effective nearest neighbor DM interaction  of $11.6$~meV can be fitted, which is in good agreement
with the corresponding value of $9.2$~meV reported in Ref. \cite{PhysRevLett.101.027201} using our convention for the exchange interactions, see above. The spin-spiral energies based on the spin model Eq.~(\ref{hamiltonian}) with the tensorial interactions from the RTM but excluding the anisotropy terms are also presented in Fig. \ref{etot-mnw001}(a). In accordance with the spin-spiral calculations this also prefers a right-handed spin-spiral, nevertheless with a somewhat smaller period of $1.6$~nm. 

We can account for the magnetic anisotropy as explained in Section~II.~C.  Within the framework of magnetic force theorem, we found an out-of plane anisotropy with value of $K=4.15$~meV. This implies that the energy of the spin spirals should be shifted upwards with respect to the energy of the FM state by $K/2=2.08$~meV, close to the value of $1.8$~meV reported in Ref. \onlinecite{PhysRevLett.101.027201}. Since this energy shift  is much smaller than the energy gain of the spin spiral due to isotropic and DM interactions ($>$15~meV, see Fig.~\ref{etot-mnw001}(a)), this spin-spiral state remains lower in energy than the FM state.

\subsection{Co monolayer on Pt(111)}
Next we performed spin-spiral calculations for a Co monolayer on Pt(111). The self-consistently treated layer structure considered in the SKKR method consisted of five Pt atomic layers, one Co monolayer and three layers of empty spheres between a semi-infinite Pt substrate and a semi-infinite vacuum region. For modelling the geometry of the system we used the in-plane lattice constant of Pt(111), $a_{2\rm D}=2.774$ \AA , fcc growth was assumed for the Pt layers and hcp stacking was used for the Co monolayer. The distance between the atomic layers were optimized in terms of VASP calculations. Relative to the interlayer distance in bulk Pt, for the Co monolayer we found an inward relaxation of $11$\% .

\begin{figure} [htb!]
\centering
\includegraphics[width=1.00\columnwidth]{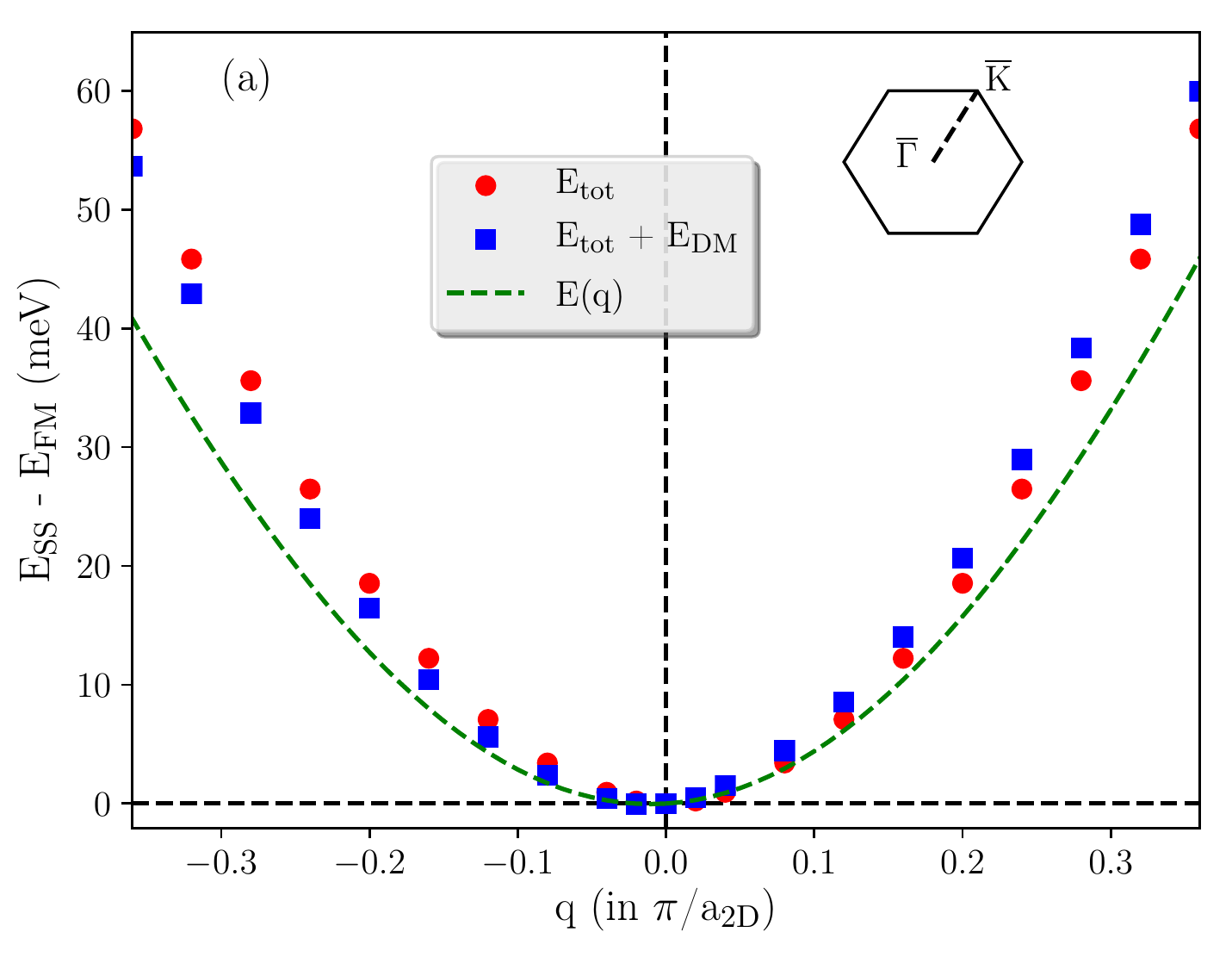}\\
\includegraphics[width=1.00\columnwidth]{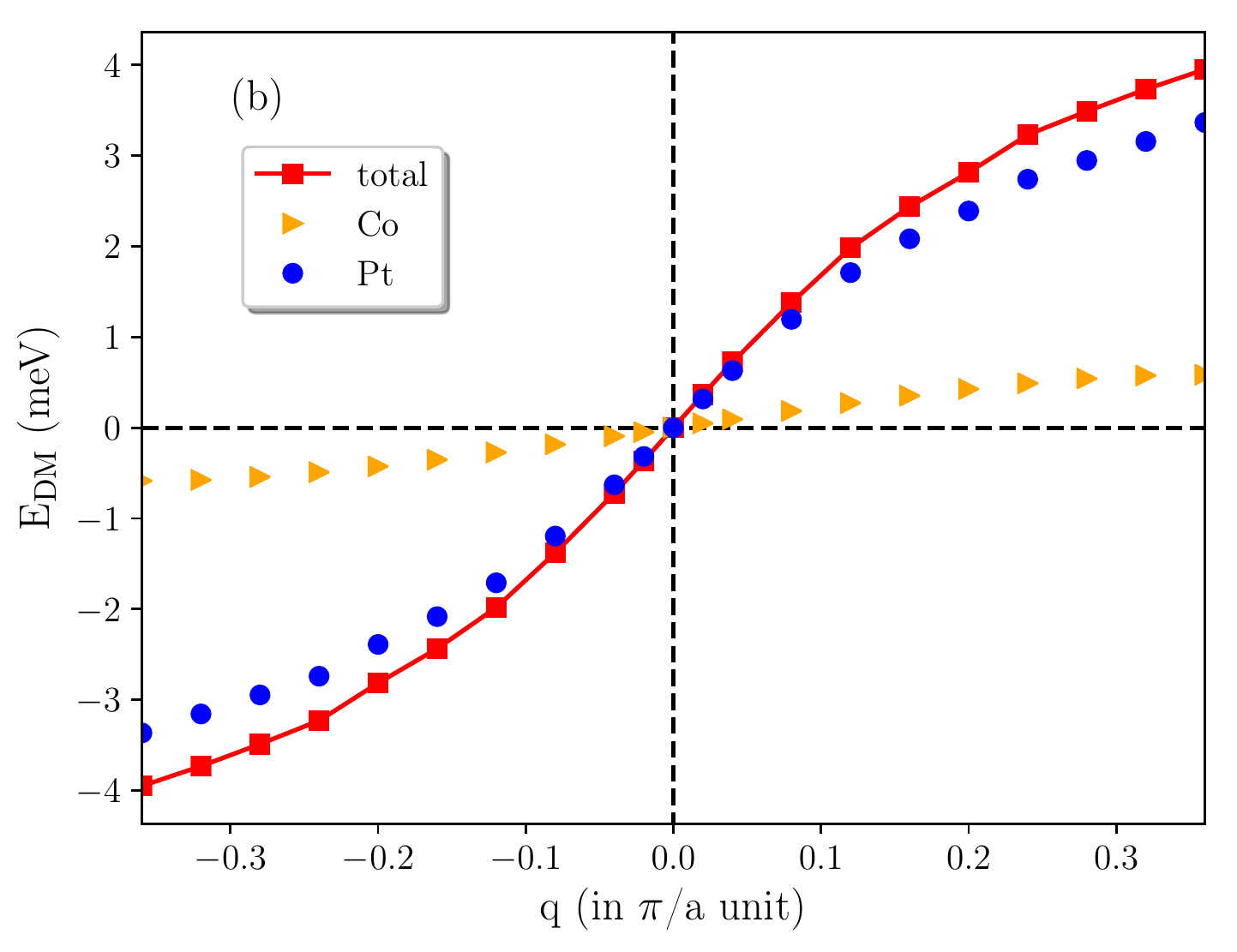}
\caption{(a) Calculated self-consistent non-relativistic total energies relative to the ferromagnetic state, E$_{\rm tot}$ and as corrected with the DM energy, E$_{\rm tot}$+E$_{\rm DM}$ for homogeneous cycloidal spin spirals propagating along the $\overline{\Gamma \rm K}$ direction in the 2D Brillouin zone (see inset) for a Co monolayer on Pt(111) with hcp stacking. The dashed line, E(q) represents the spin-spiral energy determined from a spin model containing isotropic and DM interactions between the Co atoms calculated from the RTM. (b) DM energy of the spin spirals with a decomposition into Co and Pt  contributions.}
\label{etot-copt111}
\end{figure}

Considering homogeneous flat spin spirals rotating in the $xz$ plane (the axis $x$ denoting an in-plane nearest neighbor direction), we calculated the total energy for ${\bf q}$-vectors along the $\overline{\Gamma \rm K}$ direction in the 2D Brillouin zone. The magnetic moment of the Co atoms proved to be fairly independent of $q$ with a value of $2.13 \, \mu_{\rm B}$  and the Pt layers showed to have induced moments changing according to the spin-spiral wave vector. In Fig.~\ref{etot-copt111}(a) the calculated non-relativistic and relativistic dispersion of the spin spirals are shown, the latter one obtained by adding the DM term Eq.~\eqref{DOmega} to the non-relativistic spin spiral energy. The spin-spiral energy calculated from the spin model Eq.~\eqref{hamiltonian} with only isotropic and DM interactions determined by RTM is also presented in Fig.~\ref{etot-copt111}(a).

\begin{table}[b!]
\caption{Isotropic nearest-neighbor interaction  $J_{1}$ and next nearest-neighbor interaction $J_{2}$  between the Co moments, as well as the effective nearest-neighbor interaction $J$, the spin stiffness  $J_{\rm eff}$ and the micromagnetic exchange parameter $\mathcal{J}$ for  Co/Pt(111) with hcp stacking of the Co monolayer obtained from non-relativistic self-consistent spin-spiral calculations and from the RTM.}
\label{co-jeff-table}
\centering
\begin{ruledtabular}
\begin{tabular}{l c c c c c}
    method             &  $J_{1}$ & $J_{2}$  & $J$    &  $J_{\rm eff}$                      &  $\mathcal{J}$  \\
                           & (meV)    & (meV)     &(meV)& (meV$\cdot$ \AA$^{2}$)  & (pJ/m)                \\
\hline
   spin spiral        &        58.81        &       2.08         & 65.30       &                   376.87                                     &                47.80                    \\
    RTM              &        44.43        &       2.41         & 50.15      &                    289.45                                    &               36.71                    \\
\end{tabular}
\end{ruledtabular}
\end{table}

Within the investigated range of $q$, the non-relativistic spin-spiral dispersion turned out to be fairly parabolic, thus a fit to a Heisenberg spin model allowed us to determine the first two nearest-neighbor interactions between the Co moments, $J_1$ and $J_2$. We also calculated the  effective and micromagnetic isotropic parameters defined in Eqs.~\eqref{effpars1}, \eqref{micromagnpars} and \eqref{effNNpars}, $J_{\rm eff}$, $\mathcal{J}$ and $J$, respectively. These parameters are listed in Table~\ref{co-jeff-table}. The interactions derived from the RTM were previously reported in Ref.~\cite{PhysRevB.94.214422} for hcp stacking of the Co monolayer. For comparison, the corresponding values for $J_1$ and $J_2$, as well as for the effective and micromagnetic parameters are also presented in Table~\ref{co-jeff-table}.
We find that the self-consistent spin-spiral calculations give a NN  isotropic coupling and effective parameters by about 30$\,$\% larger  than the respective parameters from the torque method. This difference can be attributed to the ferromagnetic coupling between the Co moments and the induced moments of Pt  that are included in the self-consistent spin-spiral calculations, but not taken into account in the Co-Co interactions obtained from the RTM.

By using Eq.~\eqref{DOmega}  we calculated the DM contribution to the spin-spiral energy and presented it in Fig. \ref{etot-copt111}(b). Due to the large SOC of Pt, the DM energy mainly originates from the topmost Pt layer, while the Co layer has a much smaller contribution.
From the slope of the E$_{\rm DM}$ curve at $q=0$ we determined the spiralization $D_{\rm eff}$, the microscopic DM parameter $\mathcal{D}$ from Eq.~\eqref{micromagnpars} and the effective nearest neighbor DM coupling $D$ from Eq.~\eqref{effNNpars}.  We also derived these parameters based on the previously reported in-plane DM interactions calculated in terms of the RTM \cite{PhysRevB.94.214422} and summarized them  in Table~\ref{co-deff-table}. Most likely again due to the strong interaction between the Co and Pt moments,  the effective DM parameters turned out to be by about 50~\% larger in case of the spin-spiral calculations as compared to the torque method and in both calculations the rotational direction of the DM vectors is left-handed (counterclockwise) in agreement with other theoretical results~\cite{PhysRevLett.115.267210,PhysRevB.94.214422,Freimuth2014,Dupe2014}.
 Due to the large ferromagnetic isotropic coupling, the energy gain from the formation of spin-spiral states is negligible and the easy-axis MAE of $0.57$~meV  (see in Ref. \cite{PhysRevB.94.214422} ) clearly stabilizes an out-of-plane ferromagnetic ground state. 

\begin{table}[htb!]
\caption{Effective nearest-neighbor DM interaction $D$, spiralization $D_{\rm eff}$ and micromagnetic DM parameter $\mathcal{D}$ for a Co monolayer deposited on Pt(111)  with hcp stacking  obtained from fitting to $E_{\rm DM}$ in Fig.~\ref{etot-copt111}(b)  and directly from the RTM.}
\label{co-deff-table}
\centering
\begin{ruledtabular}
\begin{tabular}{l  c c c}
   method            &   $D$  (meV)   &  $D_{\rm eff}$   (meV$\cdot$ \AA)                      &  $\mathcal{D}$ (mJ/m$^{2}$)  \\
\hline
   spin spiral        &      2.84          &       11.82         &  14.99       \\
    RTM              &      1.84       &          7.65         &     9.71
\end{tabular}
\end{ruledtabular}
\end{table}

\subsection{Co/Pt(111) capped by a Ru overlayer}
In Ref.~\onlinecite{PhysRevB.97.134405} we showed that the DM interaction is decreased  when Co/Pt(111) is capped by a $5d$ monolayer. 
In this Section we examine the effect of the $4d$ Ru overlayer on top of Co/Pt(111) which was recently claimed to induce enhanced interfacial DMI \cite{PhysRevMaterials.3.041401}. The thin film system was modelled by four monolayers of Pt, one Co monolayer, one Ru monolayer and three monolayers of empty sheres between a semi-infinite Pt substrate and a semi-infinite vacuum region. We used the in-plane lattice constant of the Pt(111) surface, $a_{2\rm D}=2.774$ \AA. From VASP calculations we have found an inward relaxation for the Co monolayer and for the Ru overlayer of $-8$\% and $-15$\% relative to the interlayer distance in bulk Pt, respectively. We assumed fcc growth for the Pt layers fcc, while hcp stacking for the Co and Ru monolayers.  

\begin{figure} [htb!]
\centering
\includegraphics[width=1.0\columnwidth]{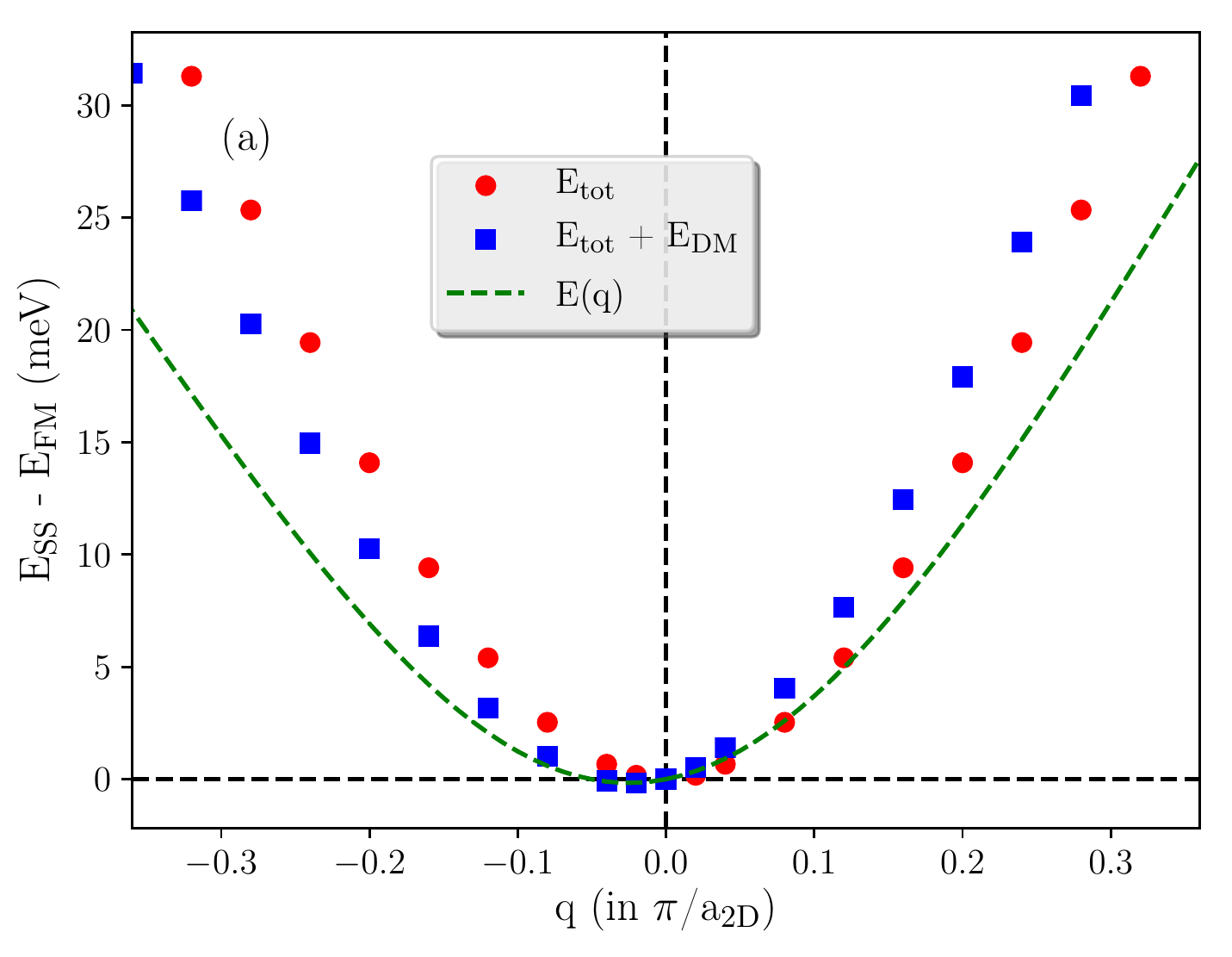}\\
\includegraphics[width=1.0\columnwidth]{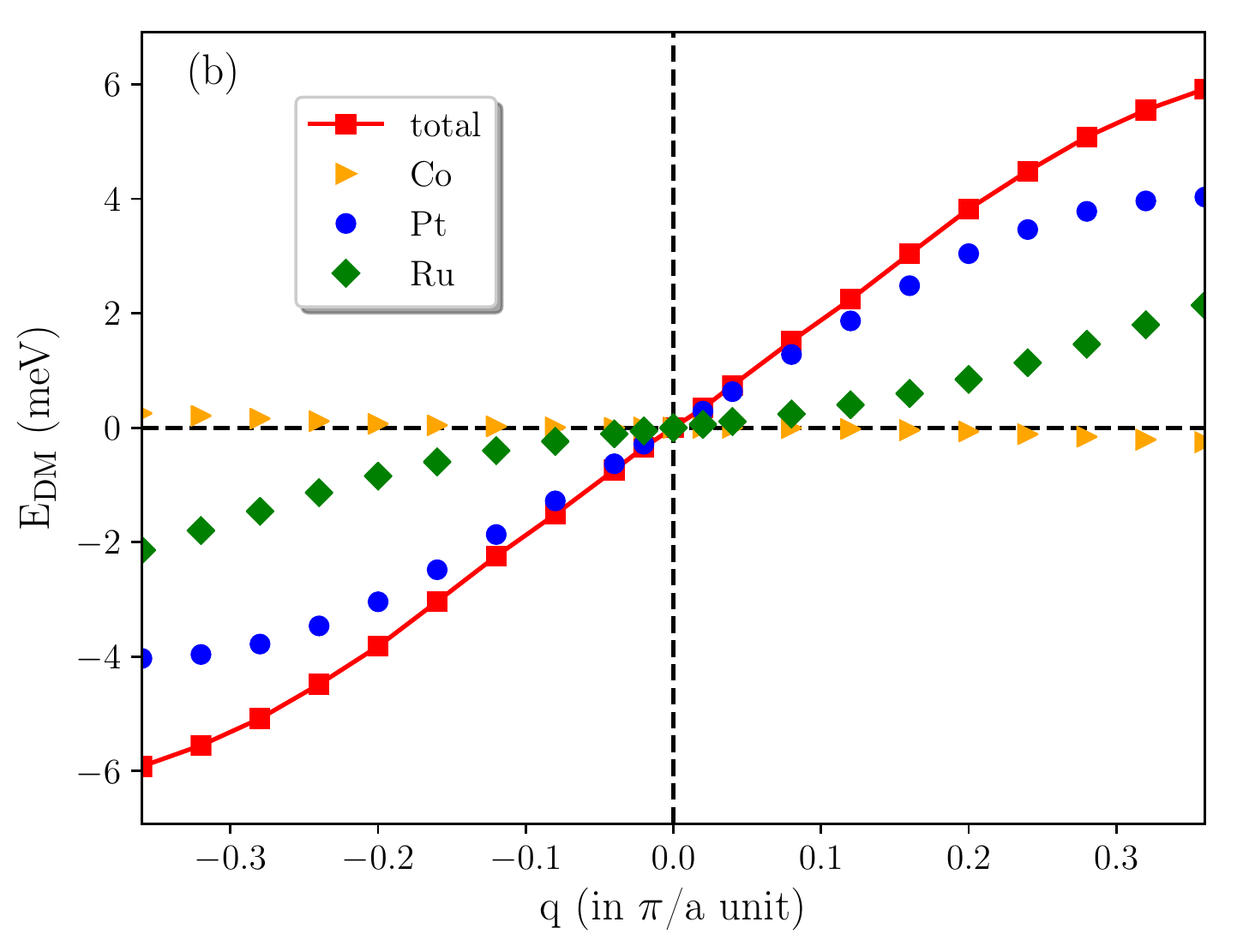}
\caption{(a) Calculated non-relativistic self-consistent spin-spiral total energies relative to the ferromagnetic state, E$_{\rm  tot}$ and the dispersion including the DM energy, E$_{\rm tot}$+E$_{\rm DM}$ along the $\overline{\Gamma \rm K}$ direction in the 2D Brillouin zone for a Co monolayer on Pt(111) capped by a Ru overlayer. The dashed line, E(q) represents the spin spiral energy determined from isotropic and DM interactions derived from the RTM.  (b) DM energy and its contributions from the Pt layers, from the Co and the Ru monolayers.}
\label{etot-rucopt111}
\end{figure}

Figure \ref{etot-rucopt111}(a) shows the spin-spiral dispersion as a function of the spin-spiral wave vector along the $\overline{\Gamma \rm K}$ direction for the Ru/Co/Pt(111) system. The magnitude of the Co magnetic moment of $1.90\mu_{B}$ was fairly the same for any $q$, while the Ru and Pt layers exhibited induced moments decreasing in size with increasing $q$.
Similar to the uncapped Co/Pt(111) monolayer, the non-relativistic spin-spiral dispersion indicates the preference of a ferromagnetic order. From a parabolic fit of the non-relativistic dispersion we obtained an effective isotropic nearest-neighbor interaction of  $J_{\rm SS} = 44.43$~meV, implying that the Ru overlayer decreased the FM coupling between the Co atoms as compared to the uncapped Co/Pt(111) system (compare with the value of 65.30~meV in Table~ \ref{co-jeff-table}). From the RTM calculations we determined a value of $J_{\rm RTM} = 35.72$~meV, which also reflects the reduced isotropic Co-Co interaction. 

The calculated Dzyaloshinskii-Moriya energy and its layer-wise composition is presented in Fig.~\ref{etot-rucopt111}(b). Apparently, the Ru overlayer significantly contributes to E$_{\rm DM}$  for larger $q$ values and it has the same sign as the contribution of the Pt layers, which is slightly enhanced as compared to the case of uncapped Co/Pt(111).  In spite that the Co monolayer has now a remarkably reduced contribution, the Ru overlayer overall increase the DM energy.  This is reflected also in the effective NN DM interaction of $D_{\rm SS} = 3.87$~meV which is by about 1~meV larger than for the uncapped system, see in Table~\ref{co-deff-table}. From the RTM method we also found a larger effective DM parameter, $D_{\rm RTM} = 3.14$~meV as compared to $D_{\rm RTM} = 1.84$~meV for Co/Pt(111). The rotational direction of the in-plane DM vectors is left-handed as in the Co/Pt(111) system. 
The Ru overlayer also drastically modified the magnetic anisotropy, because the preferred magnetization direction became in-plane, while for the uncapped system we obtained an out-of plane magnetization. The obtained in-plane anisotropy energy, $E_x-E_z=-0.56$ meV, is larger in magnitude than the very small energy gain from the formation of a spin-spiral state, indicating that the magnetic ground state is in-plane ferromagnetic. Note that in Ref.~\onlinecite{PhysRevMaterials.3.041401} a perpendicular magnetic anisotropy has been observed for the Co/Pt/Ru multilayers, but this does not contradict our present result, which refers to an overlayer system with a free surface.  

\subsection{Pt/(FeCo)/Ir multilayer systems}
Finally, we investigated periodic  Pt/(FeCo)/Ir  superlattices to model the [Ir(10$\,$\AA )/Fe(0-6$\,$\AA )/Co(4-6$\,$\AA )/Pt(10$\,$\AA )]$_{20}$ multilayers  in which room-temperature magnetic skyrmions have been found recently \citep{Soumyanarayanan2017}. In the calculations we considered an Fe and a Co monolayer in both possible sequences,  sandwiched between an Ir and a Pt bilayer, and repeated this unit periodically along the direction normal to the planes. As what follows we will label these multilayers by Pt/Co/Fe/Ir and Pt/Fe/Co/Ir.  For the hexagonal layers we used the in-plane lattice constant of Ir(111),  $a_{2\rm D}=2.714$ \AA . The Pt, Ir and Fe monolayers were stacked in fcc geometry. For the Co monolayer both hcp and fcc stackings were considered, however, the self-consistent spin-spiral calculations were performed only for the fcc stacking of the Co layer. 
The interlayer distances were optimized from VASP calculations, where we found that the interlayer distances were independent on the stacking of the Co layer. The calculated interlayer distances are summarized in Table \ref{dist_table}. These interlayer distances were used in the self-consistent spin-spiral SKKR and RTM calculations and the Wigner-Seitz radii of the atomic spheres in the Co, Fe, Pt and Ir layers were modified according to the relaxations. 

\begin{table}[htb!]
\caption{Interlayer distances for the two considered multilayer structures, Pt/Fe/Co/Ir and Pt/Co/Fe/Ir, after structural relaxations by using the VASP code. All distances are given in units of \AA .}
\label{dist_table}
\centering
\begin{ruledtabular}
\begin{tabular}{l r r r r r}
        &    d$_{\rm Pt - \rm Pt}$  &  d$_{\rm Pt - \rm Fe}$  & d$_{\rm Fe - \rm Co}$  & d$_{\rm Co- \rm Pt}$ &   d$_{\rm Ir - \rm Ir}$  \\
\hline
Pt/Fe/Co/Ir    &       2.56                  &            2.14                            &        2.02          &        2.10   &  2.21    \\

        &    d$_{\rm Pt - \rm Pt}$  &  d$_{\rm Pt - \rm Co}$  & d$_{\rm Co - \rm Fe}$  & d$_{\rm Fe- \rm Pt}$ &   d$_{\rm Ir - \rm Ir}$  \\
\hline
Pt/Co/Fe/Ir    &        2.54                 &          2.15                            &           1.99        &       2.12     &  2.22 \\

\end{tabular}
\end{ruledtabular}
\end{table}

\begin{figure} [htb!]
\centering
\includegraphics[width=1.0\columnwidth]{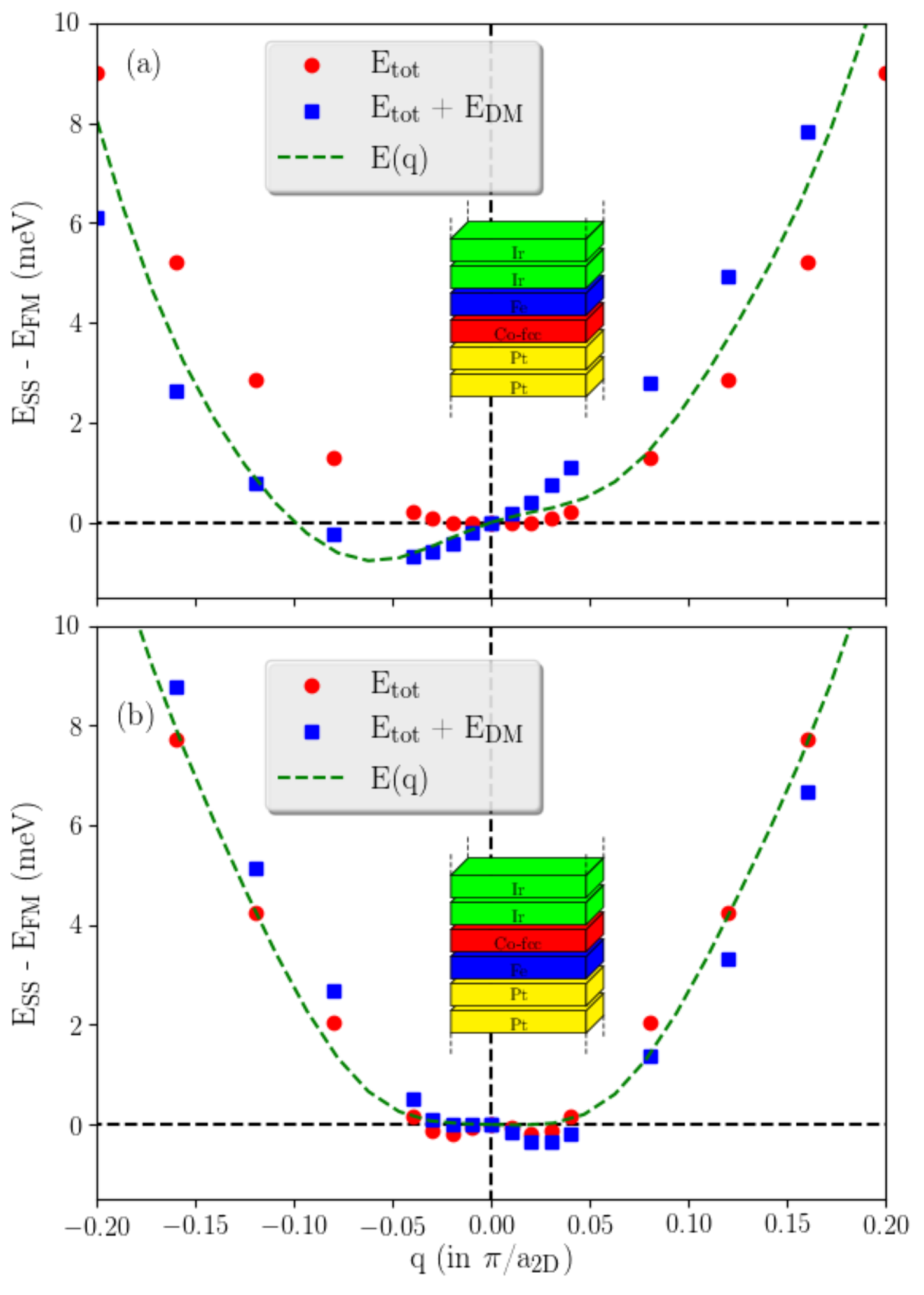}
\caption{Calculated self-consistent non-relativistic total energies of out-of-plane cycloidal spin spirals, E$_{\rm tot}$, and as corrected with the DM energy, E$_{\rm tot}+ \rm E_{\rm DM}$, along the $\overline{\Gamma \rm K}$ direction of the 2D Brillouin zone in case of (a) Pt/Co/Fe/Ir  and (b) Pt/Fe/Co/Ir (b) multilayers with fcc stacking of the Co monolayer. The dashed line, E(q) depicts the energy of the spin spirals calculated by using a spin model containing isotropic and DM interactions determined from the RTM.   The insets illustrate  the layer sequence in a unit of the multilayers.}
\label{etot-multilayer}
\end{figure}

The calculated spin-spiral dispersions for the Pt/Fe/Co/Ir and Pt/Co/Fe/Ir multilayer structures are shown in Fig.~\ref{etot-multilayer}  for a propagation direction along the $\overline{\Gamma \rm K}$ line in the 2D Brillouin zone. 
The fairly stable moments of the Fe and Co were $m_{\rm Fe}=2.84\,$$\mu_{\rm B}$ and $m_{\rm Co}=1.59\,$$\mu_{\rm B}$  for Pt/Fe/Co/Ir, and $m_{\rm Co}=1.83\,$$\mu_{\rm B}$ and $m_{\rm Fe}=2.43\,$$\mu_{\rm B}$ for Pt/Co/Fe/Ir. The non-relativistic total energy dispersion curve shows a faster increase for Pt/Fe/Co/Pt than for Pt/Co/Fe/Pt, while in both cases a shallow minimum can be found at small $q$, which indicates the appearance of frustrated isotropic interactions.  

\begin{figure} [htb!]
\centering
\includegraphics[width=1.0\columnwidth]{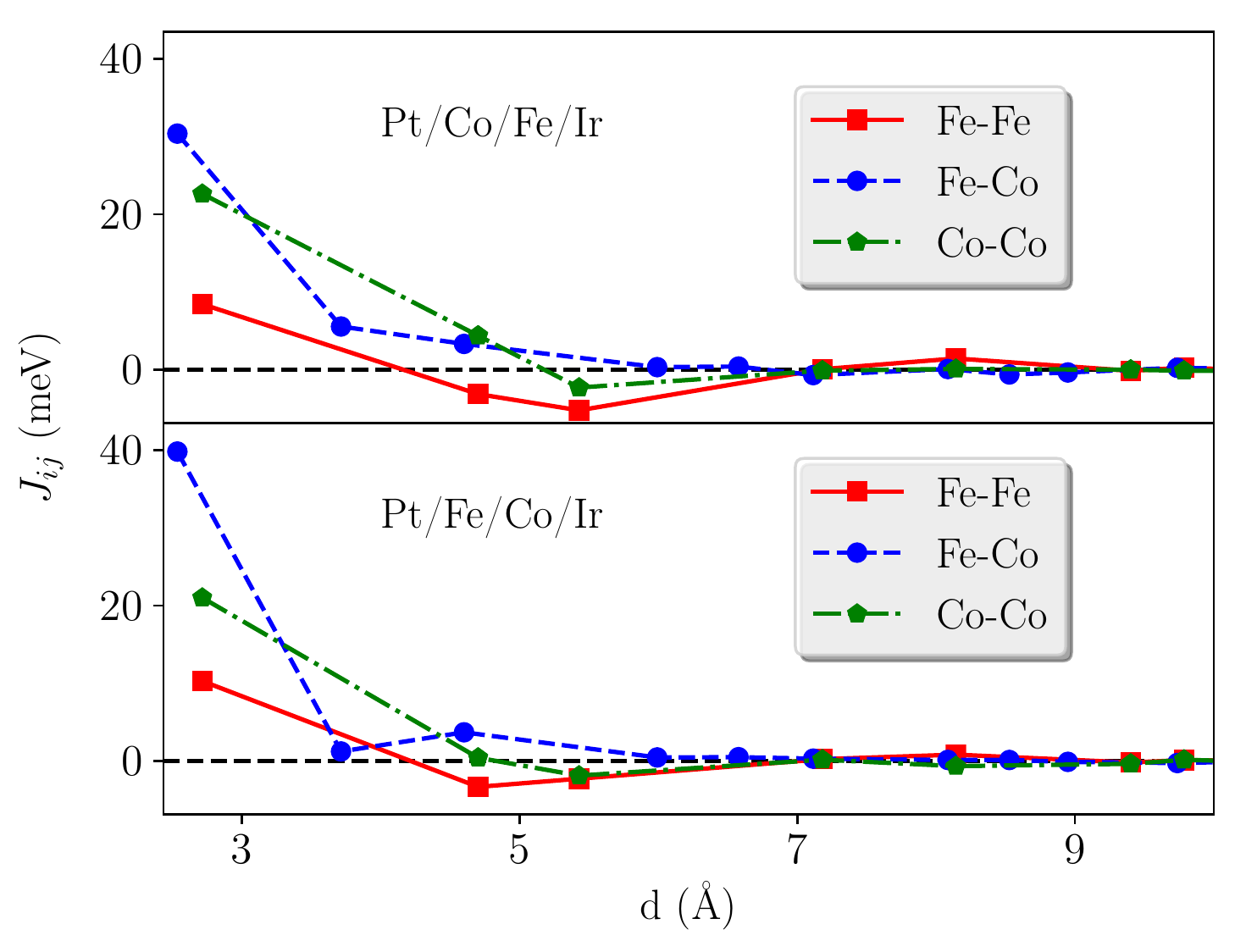}
\caption{Fe-Fe, Fe-Co and Co-Co isotropic exchange interactions $J_{ij}$ as a function of the interatomic distance $d$ for the Pt/Fe/Co/Ir and Pt/Co/Fe/Ir multilayer systems in case of fcc-stacking of the Co monolayer determined from the relativistic torque method. The interactions only within a bilayer of  CoFe or FeCo are shown.}
\label{jiso-multilayer}
\end{figure}

We calculated the isotropic exchange interactions for the Fe-Fe, Fe-Co and Co-Co pairs using the relativistic torque method and plotted them in Fig.~\ref{jiso-multilayer} as a function of the interatomic distances in case of an fcc-stacking of the Co layer. Note that interactions are presented only within a bilayer of CoFe or FeCo, since the interactions between the bilayers are negligible.  In both multilayer structures the ferromagnetic NN  Fe-Fe, Fe-Co and Co-Co interactions are dominating. The NN Fe-Co interaction in Pt/Fe/Co/Ir is clearly enhanced as compared to Pt/Co/Fe/Ir, being the main reason for the steeper spin-spiral energy dispersion for Pt/Fe/Co/Ir seen in Fig.~\ref{etot-multilayer}. While the Fe-Co interactions remain ferromagnetic for larger distances, the second and third NN  Fe-Fe interactions, as well as the third NN Co-Co interactions are antiferromagnetic, which gives rise to frustration and to the stabilization of long wavelength spin spirals as seen in Fig.~\ref{etot-multilayer}.  

\begin{figure} [htb!]
\centering
\includegraphics[width=1.0\columnwidth]{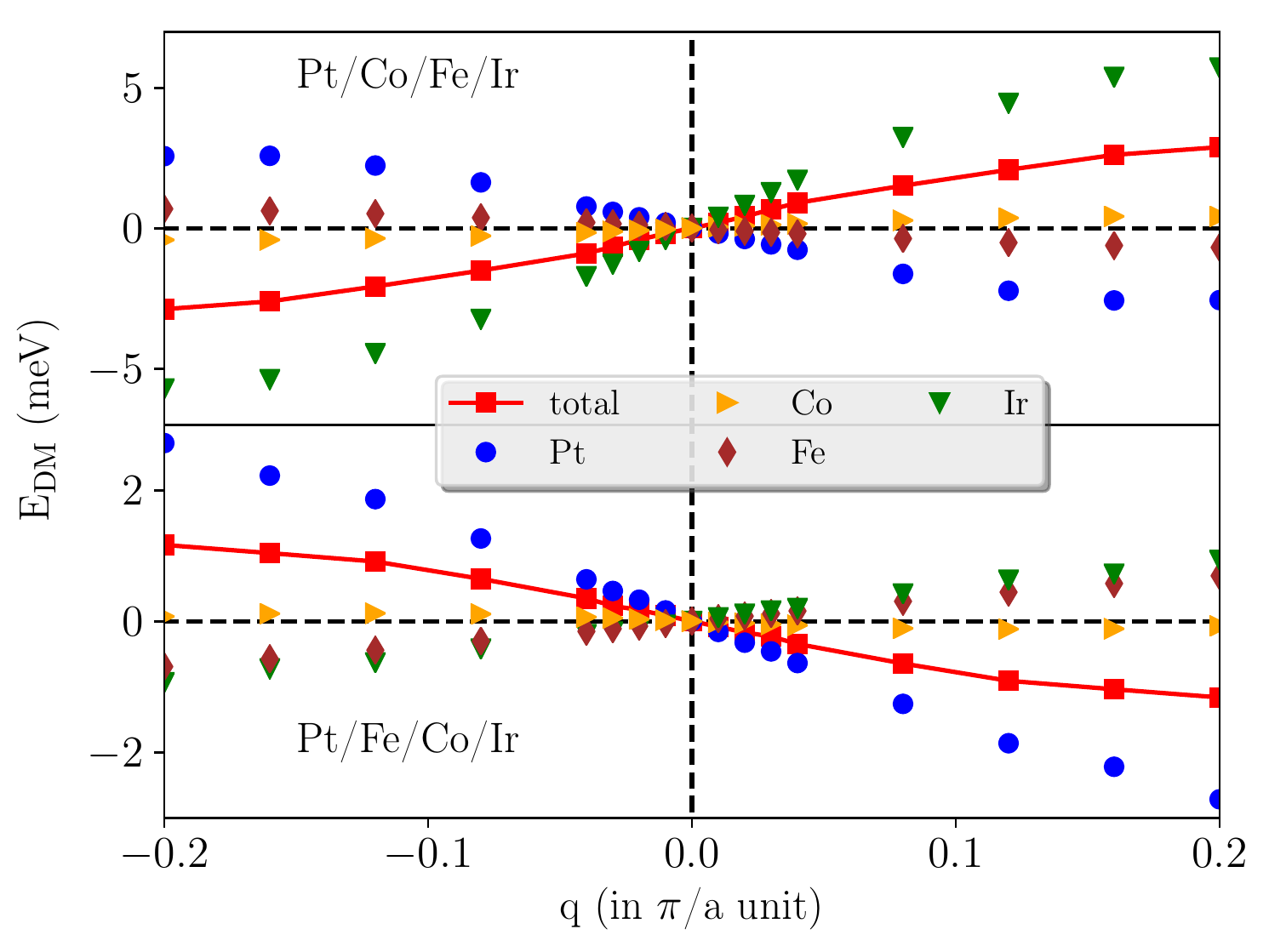}
\caption{DM energy of out-of-plane cycloidal spin spirals, E$_{\rm DM}$, propagating along the $\overline{\Gamma \rm K}$ direction of the 2D Brillouin zone  for the Pt/Fe/Co/Ir and Pt/Co/Fe/Ir multilayer systems with fcc stacking of the Co monolayer as calculated from Eq.~\eqref{DOmega}. The contributions of the Ir, Pt, Fe and Co layers are also shown. }
\label{dm-multilayer}
\end{figure}

The DM energy as calculated from Eq.~\eqref{DOmega} and its resolution into layer-wise contributions is shown in Fig.~\ref{dm-multilayer} for  an fcc stacking of the Co monolayer. In case of Pt/Fe/Co/Ir the Pt layers have a dominating contribution to E$_{\rm DM}$, while the contribution of the Ir layers is about ten times smaller  and, surprisingly, it is similar in magnitude as the contribution of the Fe layer. Moreover, the contribution of Pt is different in sign as those of Ir and Fe, and E$_{\rm DM}$ prefers a right-handed spin-spiral state.  In case of Pt/Co/Fe/Ir structure the contribution of the Pt  layers is practically unchanged, but the contribution of the Ir layers enhances in size by a factor of about five and becomes dominant in  E$_{\rm DM}$. The contribution of Fe considerably decreases and also reverses sign. Note that in both cases the DM energy related to Co is negligible. As a result, E$_{\rm DM}$  for Pt/Co/Fe/Ir reverses sign as compared with Pt/Fe/Co/Ir, thus it favors left-handed spin spirals.  

From the slope of E$_{\rm DM}$ at $q=0$ we obtained the spiralization $D_{\rm eff}$ and the effective nearest-neighbor DM interaction and presented them in Table \ref{deff_FeCo_table}  with the corresponding parameters calculated in terms of the RTM for both the fcc and hcp stacking of the Co monolayer. In case of fcc stacking of the Co monolayer, by using both computational methods the sign of the effective DM parameters is different for the two multilayer systems indicating that the sense of rotation of the DM vectors depends on the sequence of the magnetic layers. On the contrary, for hcp stacking of the Co monolayer, the sign of the DMI does not change and similar to the case of together fcc stacking of Co, the effective DM coupling is three times larger for Pt/Co/Fe/Ir than for Pt/Fe/Co/Ir. 

\begin{table}[htb!]
\caption{Effective nearest neighbor DM interaction $D$ and spiralization $D_{\rm eff}$ of the CoFe bilayers in the Pt/Fe/Co/Ir and Pt/Co/Fe/Ir multilayers obtained from spin-spiral calculations and from the relativistic torque method with different stackings of the Co layer.}
\label{deff_FeCo_table}
\centering
\begin{ruledtabular}
\begin{tabular}{c c c c c}
                     &stacking              & method   & $D$ (meV)  &  $D_{\rm eff}$ (meV$\cdot$ \AA) \\
\hline
                    &Co-fcc    &  spin spiral      & -1.78 & 7.25   \\
Pt/Fe/Co/Ir & Co-fcc               &   RTM          &          -0.45  & -1.83 \\
                     &Co-hcp               &   RTM          &        1.11  & 4.52  \\
\hline
                    &Co-fcc    &  spin spiral     &    4.57   & -18.60             \\
Pt/Co/Fe/Ir & Co-fcc                  &  RTM           &        2.66    & 10.83     \\
                     & Co-hcp                  &  RTM          &        3.28      & 13.35    \\
\end{tabular}
\end{ruledtabular}
\end{table}

By using the magnetic force theorem we also determined the magnetic anisotropy energy of the multilayer systems for both kinds of Co stacking. We have found that the Pt/Fe/Co/Ir multilayer shows perpendicular magnetic anisotropy in case of both the fcc and the hcp stacking of Co, with $\Delta E=E_x - E_z=0.76$~meV and $0.36$~meV, respectively. In case of the Pt/Co/Fe/Ir multilayer with Co-fcc stacking we obtained an in-plane anisotropy with $\Delta E=-0.64$~meV, while for Co-hcp stacking an easy-axis anisotropy with $\Delta E =0.59$~meV. It is known, however, that the magnetic anisotropy of multilayers is very sensitive to the growth conditions \cite{DENBROEDER1991562, PhysRevB.41.11919, Wu_2012} and, similarly as in Ref.~\onlinecite{Soumyanarayanan2017}, our calculated values of the MAE are much larger than the experimental values. Due to the frustrated isotropic couplings between the Fe and Co atoms and due to the large DM interaction, skyrmionic states can then likely be stabilized in Pt/Co/Fe/Ir multilayers. 

\section{Summary and conclusions}
\label{sec:summary}
In summary, by using the multiple scattering Green's function technique we presented a theoretical approach to calculate the electronic structure of layered systems with spiral magnetic structure. For layered systems the propagation direction of the spin-spiral is restricted to the plane of the layers, while an arbitrary rotational direction can be chosen. The non-relativistic allows for self-consistent calculations, from which the total energy of the system can be obtained as a function of the spin-wave vector. We employed a first-order perturbation technique to include the effect of spin-orbit coupling in the calculations. A particular advantage of this approach is that the energy related to the Dzyaloshinskii-Moriya interactions can be resolved into layer-wise contributions. 

We performed ab initio calculations for ultrathin films and multilayers and demonstrated that the newly developed method gives an accurate access to the magnetism of these systems. We found that the magnetic ground state of a Mn monolayer on W(001) is a right-handed spin-spiral in good agreement with the experiment and calculations in Ref.~\onlinecite{PhysRevLett.101.027201}. From the spin-spiral dispersion we derived spin-model parameters, which compared well with the couplings calculated by the relativistic torque method. We investigated the spin-spiral states of a Co monolayer on Pt(111) with hcp stacking of the Co monolayer and concluded that the ground state of the system is ferromagnetic, similar to previous calculations  \cite{Dupe2014, PhysRevB.99.214426}. The nearest-neighbor isotropic exchange interaction determined from the non-relativistic spin-spiral dispersion was found significantly larger than the corresponding interaction from the torque method which can be attributed to the effect of the induced moments of the Pt atoms, included inherently in the spin-spiral calculations. For the Co monolayer on Pt(111) capped by a Ru overlayer we showed that the isotropic coupling between the Co atoms is reduced, while the interfacial DMI was increased. These results correlate well with recent experiments on Pt/Co/Ru superlattices \cite{PhysRevMaterials.3.041401}. We also investigated Pt/Fe/Co/Ir and Pt/Co/Fe/Ir multilayers and found that the non-relativistic dispersion implies the appearence of spin-spiral states with large wavelength due to the frustrated couplings between the Fe and Co atoms. Remarkably, the rotational sense of the DMI was opposite for the two multilayers, which could be attributed to the largely enhanced contribution of the Ir layer in the Pt/Co/Fe/Ir  multilayer. Moreover, we found that the effective DMI is three times larger for Pt/Co/Fe/Ir than in case of Pt/Fe/Co/Ir independently on the stacking of the Co layer. 

Our results obtained from the presented spin-spiral approach provide thus a theoretical support to the fine-tuning of the magnetic and non-magnetic layers in multilayer structures with the purpose of manipulating the interfacial DMI and designing new building elements for spintronics applications \cite{Soumyanarayanan2017}. An obvious possibility to proceed on this way is to consider disordered alloys in these structures that is easily feasible within the Green's function technique in terms of the coherent potential approximation \cite{PhysRevB.83.144401}.

\begin{acknowledgments}
The authors are grateful for illuminating discussions with Manuel Dos Santos Dias and Laszlo Udvardi, and acknowledge NIIF and PRACE for awarding access to the resources based in Hungary at Debrecen and Szeged, and in Germany at J\"ulich. This work was supported by the National Research, Development and Innovation Office of Hungary under projects No. PD120917 and K115575, as well as by the BME Nanotechnology FIKP grant of EMMI (BME FIKP-NAT).
\end{acknowledgments}

%
%% References with BibTex database
\bibliographystyle{apsrev4-1}
\bibliography{references}

\begin{thebibliography}{63}%
\makeatletter
\providecommand \@ifxundefined [1]{%
 \@ifx{#1\undefined}
}%
\providecommand \@ifnum [1]{%
 \ifnum #1\expandafter \@firstoftwo
 \else \expandafter \@secondoftwo
 \fi
}%
\providecommand \@ifx [1]{%
 \ifx #1\expandafter \@firstoftwo
 \else \expandafter \@secondoftwo
 \fi
}%
\providecommand \natexlab [1]{#1}%
\providecommand \enquote  [1]{``#1''}%
\providecommand \bibnamefont  [1]{#1}%
\providecommand \bibfnamefont [1]{#1}%
\providecommand \citenamefont [1]{#1}%
\providecommand \href@noop [0]{\@secondoftwo}%
\providecommand \href [0]{\begingroup \@sanitize@url \@href}%
\providecommand \@href[1]{\@@startlink{#1}\@@href}%
\providecommand \@@href[1]{\endgroup#1\@@endlink}%
\providecommand \@sanitize@url [0]{\catcode `\\12\catcode `\$12\catcode
  `\&12\catcode `\#12\catcode `\^12\catcode `\_12\catcode `\%12\relax}%
\providecommand \@@startlink[1]{}%
\providecommand \@@endlink[0]{}%
\providecommand \url  [0]{\begingroup\@sanitize@url \@url }%
\providecommand \@url [1]{\endgroup\@href {#1}{\urlprefix }}%
\providecommand \urlprefix  [0]{URL }%
\providecommand \Eprint [0]{\href }%
\providecommand \doibase [0]{http://dx.doi.org/}%
\providecommand \selectlanguage [0]{\@gobble}%
\providecommand \bibinfo  [0]{\@secondoftwo}%
\providecommand \bibfield  [0]{\@secondoftwo}%
\providecommand \translation [1]{[#1]}%
\providecommand \BibitemOpen [0]{}%
\providecommand \bibitemStop [0]{}%
\providecommand \bibitemNoStop [0]{.\EOS\space}%
\providecommand \EOS [0]{\spacefactor3000\relax}%
\providecommand \BibitemShut  [1]{\csname bibitem#1\endcsname}%
\let\auto@bib@innerbib\@empty
%</preamble>
\bibitem [{\citenamefont {Parkin}\ \emph {et~al.}(2008)\citenamefont {Parkin},
  \citenamefont {Hayashi},\ and\ \citenamefont {Thomas}}]{Parkin}%
  \BibitemOpen
  \bibfield  {author} {\bibinfo {author} {\bibfnamefont {S.~S.~P.}\
  \bibnamefont {Parkin}}, \bibinfo {author} {\bibfnamefont {M.}~\bibnamefont
  {Hayashi}}, \ and\ \bibinfo {author} {\bibfnamefont {L.}~\bibnamefont
  {Thomas}},\ }\href {\doibase 10.1126/science.1145799} {\bibfield  {journal}
  {\bibinfo  {journal} {Science}\ }\textbf {\bibinfo {volume} {320}},\ \bibinfo
  {pages} {190} (\bibinfo {year} {2008})}\BibitemShut {NoStop}%
\bibitem [{\citenamefont {Fert}\ \emph {et~al.}(2013)\citenamefont {Fert},
  \citenamefont {Cros},\ and\ \citenamefont {Sampaio}}]{Fert}%
  \BibitemOpen
  \bibfield  {author} {\bibinfo {author} {\bibfnamefont {A.}~\bibnamefont
  {Fert}}, \bibinfo {author} {\bibfnamefont {V.}~\bibnamefont {Cros}}, \ and\
  \bibinfo {author} {\bibfnamefont {J.}~\bibnamefont {Sampaio}},\ }\href@noop
  {} {\bibfield  {journal} {\bibinfo  {journal} {Nat. Nanotechnol.}\ }\textbf
  {\bibinfo {volume} {8}},\ \bibinfo {pages} {152} (\bibinfo {year}
  {2013})}\BibitemShut {NoStop}%
\bibitem [{\citenamefont {Duine}\ \emph {et~al.}(2018)\citenamefont {Duine},
  \citenamefont {Lee}, \citenamefont {Parkin},\ and\ \citenamefont
  {Stiles}}]{Duine2018}%
  \BibitemOpen
  \bibfield  {author} {\bibinfo {author} {\bibfnamefont {R.~A.}\ \bibnamefont
  {Duine}}, \bibinfo {author} {\bibfnamefont {K.-J.}\ \bibnamefont {Lee}},
  \bibinfo {author} {\bibfnamefont {S.~S.~P.}\ \bibnamefont {Parkin}}, \ and\
  \bibinfo {author} {\bibfnamefont {M.~D.}\ \bibnamefont {Stiles}},\ }\href
  {\doibase 10.1038/s41567-018-0050-y} {\bibfield  {journal} {\bibinfo
  {journal} {Nat. Phys.}\ }\textbf {\bibinfo {volume} {14}},\ \bibinfo {pages}
  {217} (\bibinfo {year} {2018})}\BibitemShut {NoStop}%
\bibitem [{\citenamefont {Bihlmayer}(2007)}]{Bihlmayer}%
  \BibitemOpen
  \bibfield  {author} {\bibinfo {author} {\bibfnamefont {G.}~\bibnamefont
  {Bihlmayer}},\ }\enquote {\bibinfo {title} {Density‐functional theory of
  magnetism},}\ in\ \href {\doibase 10.1002/9780470022184.hmm101} {\emph
  {\bibinfo {booktitle} {Handbook of Magnetism and Advanced Magnetic
  Materials}}}\ (\bibinfo  {publisher} {Wiley, New York},\ \bibinfo {year}
  {2007})\BibitemShut {NoStop}%
\bibitem [{\citenamefont {Nowak}(2007)}]{Nowak}%
  \BibitemOpen
  \bibfield  {author} {\bibinfo {author} {\bibfnamefont {U.}~\bibnamefont
  {Nowak}},\ }\enquote {\bibinfo {title} {Classical spin models},}\ in\ \href
  {\doibase 10.1002/97804470022184.hmm205} {\emph {\bibinfo {booktitle}
  {Handbook of Magnetism and Advanced Magnetic Materials}}}\ (\bibinfo
  {publisher} {Wiley, New York},\ \bibinfo {year} {2007})\BibitemShut {NoStop}%
\bibitem [{\citenamefont {von Bergmann}\ \emph {et~al.}(2014)\citenamefont {von
  Bergmann}, \citenamefont {Kubetzka}, \citenamefont {Pietzsch},\ and\
  \citenamefont {Wiesendanger}}]{Bergmann_2014}%
  \BibitemOpen
  \bibfield  {author} {\bibinfo {author} {\bibfnamefont {K.}~\bibnamefont {von
  Bergmann}}, \bibinfo {author} {\bibfnamefont {A.}~\bibnamefont {Kubetzka}},
  \bibinfo {author} {\bibfnamefont {O.}~\bibnamefont {Pietzsch}}, \ and\
  \bibinfo {author} {\bibfnamefont {R.}~\bibnamefont {Wiesendanger}},\ }\href
  {\doibase 10.1088/0953-8984/26/39/394002} {\bibfield  {journal} {\bibinfo
  {journal} {Journal of Physics: Condensed Matter}\ }\textbf {\bibinfo {volume}
  {26}},\ \bibinfo {pages} {394002} (\bibinfo {year} {2014})}\BibitemShut
  {NoStop}%
\bibitem [{\citenamefont {Sandratskii}(1986)}]{Sandratskii_PhysStatSol}%
  \BibitemOpen
  \bibfield  {author} {\bibinfo {author} {\bibfnamefont {L.~M.}\ \bibnamefont
  {Sandratskii}},\ }\href {\doibase 10.1002/pssb.2221360119} {\bibfield
  {journal} {\bibinfo  {journal} {Phys. Status Solidi B}\ }\textbf {\bibinfo
  {volume} {136}},\ \bibinfo {pages} {167} (\bibinfo {year}
  {1986})}\BibitemShut {NoStop}%
\bibitem [{\citenamefont {Sandratskii}(1991)}]{Sandratskii_1991}%
  \BibitemOpen
  \bibfield  {author} {\bibinfo {author} {\bibfnamefont {L.~M.}\ \bibnamefont
  {Sandratskii}},\ }\href {\doibase 10.1088/0953-8984/3/44/004} {\bibfield
  {journal} {\bibinfo  {journal} {J. Phys.: Condens. Matter}\ }\textbf
  {\bibinfo {volume} {3}},\ \bibinfo {pages} {8565} (\bibinfo {year}
  {1991})}\BibitemShut {NoStop}%
\bibitem [{\citenamefont {Sandratskii}\ and\ \citenamefont
  {Guletskii}(1986)}]{Sandratskii_1986}%
  \BibitemOpen
  \bibfield  {author} {\bibinfo {author} {\bibfnamefont {L.~M.}\ \bibnamefont
  {Sandratskii}}\ and\ \bibinfo {author} {\bibfnamefont {P.~G.}\ \bibnamefont
  {Guletskii}},\ }\href {\doibase 10.1088/0305-4608/16/2/002} {\bibfield
  {journal} {\bibinfo  {journal} {J. Phys. F: Met. Phys.}\ }\textbf {\bibinfo
  {volume} {16}},\ \bibinfo {pages} {L43} (\bibinfo {year} {1986})}\BibitemShut
  {NoStop}%
\bibitem [{\citenamefont {Mryasov}\ \emph {et~al.}(1991)\citenamefont
  {Mryasov}, \citenamefont {Liechtenstein}, \citenamefont {Sandratskii},\ and\
  \citenamefont {Gubanov}}]{Mryasov_1991}%
  \BibitemOpen
  \bibfield  {author} {\bibinfo {author} {\bibfnamefont {O.~N.}\ \bibnamefont
  {Mryasov}}, \bibinfo {author} {\bibfnamefont {A.~I.}\ \bibnamefont
  {Liechtenstein}}, \bibinfo {author} {\bibfnamefont {L.~M.}\ \bibnamefont
  {Sandratskii}}, \ and\ \bibinfo {author} {\bibfnamefont {V.~A.}\ \bibnamefont
  {Gubanov}},\ }\href {\doibase 10.1088/0953-8984/3/39/013} {\bibfield
  {journal} {\bibinfo  {journal} {J. Phys.: Condens. Matter}\ }\textbf
  {\bibinfo {volume} {3}},\ \bibinfo {pages} {7683} (\bibinfo {year}
  {1991})}\BibitemShut {NoStop}%
\bibitem [{\citenamefont {Sandratskii}(1998)}]{Sandratskii_AdvPhys}%
  \BibitemOpen
  \bibfield  {author} {\bibinfo {author} {\bibfnamefont {L.~M.}\ \bibnamefont
  {Sandratskii}},\ }\href {\doibase 10.1080/000187398243573} {\bibfield
  {journal} {\bibinfo  {journal} {Adv. Phys.}\ }\textbf {\bibinfo {volume}
  {47}},\ \bibinfo {pages} {91} (\bibinfo {year} {1998})}\BibitemShut {NoStop}%
\bibitem [{\citenamefont {Kurz}\ \emph {et~al.}(2004)\citenamefont {Kurz},
  \citenamefont {F\"orster}, \citenamefont {Nordstr\"om}, \citenamefont
  {Bihlmayer},\ and\ \citenamefont {Bl\"ugel}}]{PhysRevB.69.024415}%
  \BibitemOpen
  \bibfield  {author} {\bibinfo {author} {\bibfnamefont {P.}~\bibnamefont
  {Kurz}}, \bibinfo {author} {\bibfnamefont {F.}~\bibnamefont {F\"orster}},
  \bibinfo {author} {\bibfnamefont {L.}~\bibnamefont {Nordstr\"om}}, \bibinfo
  {author} {\bibfnamefont {G.}~\bibnamefont {Bihlmayer}}, \ and\ \bibinfo
  {author} {\bibfnamefont {S.}~\bibnamefont {Bl\"ugel}},\ }\href {\doibase
  10.1103/PhysRevB.69.024415} {\bibfield  {journal} {\bibinfo  {journal} {Phys.
  Rev. B}\ }\textbf {\bibinfo {volume} {69}},\ \bibinfo {pages} {024415}
  (\bibinfo {year} {2004})}\BibitemShut {NoStop}%
\bibitem [{\citenamefont {Mankovsky}\ \emph {et~al.}(2011)\citenamefont
  {Mankovsky}, \citenamefont {Fecher},\ and\ \citenamefont
  {Ebert}}]{PhysRevB.83.144401}%
  \BibitemOpen
  \bibfield  {author} {\bibinfo {author} {\bibfnamefont {S.}~\bibnamefont
  {Mankovsky}}, \bibinfo {author} {\bibfnamefont {G.~H.}\ \bibnamefont
  {Fecher}}, \ and\ \bibinfo {author} {\bibfnamefont {H.}~\bibnamefont
  {Ebert}},\ }\href {\doibase 10.1103/PhysRevB.83.144401} {\bibfield  {journal}
  {\bibinfo  {journal} {Phys. Rev. B}\ }\textbf {\bibinfo {volume} {83}},\
  \bibinfo {pages} {144401} (\bibinfo {year} {2011})}\BibitemShut {NoStop}%
\bibitem [{\citenamefont {Dzyaloshinsky}(1958)}]{Dzyaloshinsky}%
  \BibitemOpen
  \bibfield  {author} {\bibinfo {author} {\bibfnamefont {I.}~\bibnamefont
  {Dzyaloshinsky}},\ }\href@noop {} {\bibfield  {journal} {\bibinfo  {journal}
  {J. Phys. Chem. Solids}\ }\textbf {\bibinfo {volume} {4}},\ \bibinfo {pages}
  {241 } (\bibinfo {year} {1958})}\BibitemShut {NoStop}%
\bibitem [{\citenamefont {Moriya}(1960)}]{Moriya}%
  \BibitemOpen
  \bibfield  {author} {\bibinfo {author} {\bibfnamefont {T.}~\bibnamefont
  {Moriya}},\ }\href@noop {} {\bibfield  {journal} {\bibinfo  {journal} {Phys.
  Rev. Lett.}\ }\textbf {\bibinfo {volume} {4}},\ \bibinfo {pages} {228}
  (\bibinfo {year} {1960})}\BibitemShut {NoStop}%
\bibitem [{\citenamefont {Udvardi}\ \emph {et~al.}(2003)\citenamefont
  {Udvardi}, \citenamefont {Szunyogh}, \citenamefont {Palot\'{a}s},\ and\
  \citenamefont {Weinberger}}]{rtm-Udvardi}%
  \BibitemOpen
  \bibfield  {author} {\bibinfo {author} {\bibfnamefont {L.}~\bibnamefont
  {Udvardi}}, \bibinfo {author} {\bibfnamefont {L.}~\bibnamefont {Szunyogh}},
  \bibinfo {author} {\bibfnamefont {K.}~\bibnamefont {Palot\'{a}s}}, \ and\
  \bibinfo {author} {\bibfnamefont {P.}~\bibnamefont {Weinberger}},\
  }\href@noop {} {\bibfield  {journal} {\bibinfo  {journal} {Phys. Rev. B}\
  }\textbf {\bibinfo {volume} {68}},\ \bibinfo {pages} {104436} (\bibinfo
  {year} {2003})}\BibitemShut {NoStop}%
\bibitem [{\citenamefont {Ebert}\ and\ \citenamefont
  {Mankovsky}(2009)}]{rtm-Ebert}%
  \BibitemOpen
  \bibfield  {author} {\bibinfo {author} {\bibfnamefont {H.}~\bibnamefont
  {Ebert}}\ and\ \bibinfo {author} {\bibfnamefont {S.}~\bibnamefont
  {Mankovsky}},\ }\href@noop {} {\bibfield  {journal} {\bibinfo  {journal}
  {Phys. Rev. B}\ }\textbf {\bibinfo {volume} {79}},\ \bibinfo {pages} {045209}
  (\bibinfo {year} {2009})}\BibitemShut {NoStop}%
\bibitem [{\citenamefont {Katsnelson}\ \emph {et~al.}(2010)\citenamefont
  {Katsnelson}, \citenamefont {Kvashnin}, \citenamefont {Mazurenko},\ and\
  \citenamefont {Lichtenstein}}]{PhysRevB.82.100403}%
  \BibitemOpen
  \bibfield  {author} {\bibinfo {author} {\bibfnamefont {M.~I.}\ \bibnamefont
  {Katsnelson}}, \bibinfo {author} {\bibfnamefont {Y.~O.}\ \bibnamefont
  {Kvashnin}}, \bibinfo {author} {\bibfnamefont {V.~V.}\ \bibnamefont
  {Mazurenko}}, \ and\ \bibinfo {author} {\bibfnamefont {A.~I.}\ \bibnamefont
  {Lichtenstein}},\ }\href {\doibase 10.1103/PhysRevB.82.100403} {\bibfield
  {journal} {\bibinfo  {journal} {Phys. Rev. B}\ }\textbf {\bibinfo {volume}
  {82}},\ \bibinfo {pages} {100403} (\bibinfo {year} {2010})}\BibitemShut
  {NoStop}%
\bibitem [{\citenamefont {Mankovsky}\ \emph {et~al.}(2019)\citenamefont
  {Mankovsky}, \citenamefont {Polesya},\ and\ \citenamefont
  {Ebert}}]{PhysRevB.99.104427}%
  \BibitemOpen
  \bibfield  {author} {\bibinfo {author} {\bibfnamefont {S.}~\bibnamefont
  {Mankovsky}}, \bibinfo {author} {\bibfnamefont {S.}~\bibnamefont {Polesya}},
  \ and\ \bibinfo {author} {\bibfnamefont {H.}~\bibnamefont {Ebert}},\ }\href
  {\doibase 10.1103/PhysRevB.99.104427} {\bibfield  {journal} {\bibinfo
  {journal} {Phys. Rev. B}\ }\textbf {\bibinfo {volume} {99}},\ \bibinfo
  {pages} {104427} (\bibinfo {year} {2019})}\BibitemShut {NoStop}%
\bibitem [{\citenamefont {Szunyogh}\ \emph {et~al.}(2011)\citenamefont
  {Szunyogh}, \citenamefont {Udvardi}, \citenamefont {Jackson}, \citenamefont
  {Nowak},\ and\ \citenamefont {Chantrell}}]{SCE-RDLM-1}%
  \BibitemOpen
  \bibfield  {author} {\bibinfo {author} {\bibfnamefont {L.}~\bibnamefont
  {Szunyogh}}, \bibinfo {author} {\bibfnamefont {L.}~\bibnamefont {Udvardi}},
  \bibinfo {author} {\bibfnamefont {J.}~\bibnamefont {Jackson}}, \bibinfo
  {author} {\bibfnamefont {U.}~\bibnamefont {Nowak}}, \ and\ \bibinfo {author}
  {\bibfnamefont {R.}~\bibnamefont {Chantrell}},\ }\href {\doibase
  10.1103/PhysRevB.83.024401} {\bibfield  {journal} {\bibinfo  {journal} {Phys.
  Rev. B}\ }\textbf {\bibinfo {volume} {83}},\ \bibinfo {pages} {024401}
  (\bibinfo {year} {2011})}\BibitemShut {NoStop}%
\bibitem [{\citenamefont {De\'ak}\ \emph {et~al.}(2011)\citenamefont {De\'ak},
  \citenamefont {Szunyogh},\ and\ \citenamefont {Ujfalussy}}]{SCE-RDLM-2}%
  \BibitemOpen
  \bibfield  {author} {\bibinfo {author} {\bibfnamefont {A.}~\bibnamefont
  {De\'ak}}, \bibinfo {author} {\bibfnamefont {L.}~\bibnamefont {Szunyogh}}, \
  and\ \bibinfo {author} {\bibfnamefont {B.}~\bibnamefont {Ujfalussy}},\ }\href
  {\doibase 10.1103/PhysRevB.84.224413} {\bibfield  {journal} {\bibinfo
  {journal} {Phys. Rev. B}\ }\textbf {\bibinfo {volume} {84}},\ \bibinfo
  {pages} {224413} (\bibinfo {year} {2011})}\BibitemShut {NoStop}%
\bibitem [{\citenamefont {Heide}\ \emph {et~al.}(2009)\citenamefont {Heide},
  \citenamefont {Bihlmayer},\ and\ \citenamefont {Blügel}}]{HEIDE20092678}%
  \BibitemOpen
  \bibfield  {author} {\bibinfo {author} {\bibfnamefont {M.}~\bibnamefont
  {Heide}}, \bibinfo {author} {\bibfnamefont {G.}~\bibnamefont {Bihlmayer}}, \
  and\ \bibinfo {author} {\bibfnamefont {S.}~\bibnamefont {Blügel}},\ }\href
  {\doibase https://doi.org/10.1016/j.physb.2009.06.070} {\bibfield  {journal}
  {\bibinfo  {journal} {Physica B: Condens. Matter}\ }\textbf {\bibinfo
  {volume} {404}},\ \bibinfo {pages} {2678 } (\bibinfo {year}
  {2009})}\BibitemShut {NoStop}%
\bibitem [{\citenamefont {Yang}\ \emph {et~al.}(2015)\citenamefont {Yang},
  \citenamefont {Thiaville}, \citenamefont {Rohart}, \citenamefont {Fert},\
  and\ \citenamefont {Chshiev}}]{PhysRevLett.115.267210}%
  \BibitemOpen
  \bibfield  {author} {\bibinfo {author} {\bibfnamefont {H.}~\bibnamefont
  {Yang}}, \bibinfo {author} {\bibfnamefont {A.}~\bibnamefont {Thiaville}},
  \bibinfo {author} {\bibfnamefont {S.}~\bibnamefont {Rohart}}, \bibinfo
  {author} {\bibfnamefont {A.}~\bibnamefont {Fert}}, \ and\ \bibinfo {author}
  {\bibfnamefont {M.}~\bibnamefont {Chshiev}},\ }\href {\doibase
  10.1103/PhysRevLett.115.267210} {\bibfield  {journal} {\bibinfo  {journal}
  {Phys. Rev. Lett.}\ }\textbf {\bibinfo {volume} {115}},\ \bibinfo {pages}
  {267210} (\bibinfo {year} {2015})}\BibitemShut {NoStop}%
\bibitem [{\citenamefont {Sandratskii}(2017)}]{PhysRevB.96.024450}%
  \BibitemOpen
  \bibfield  {author} {\bibinfo {author} {\bibfnamefont {L.~M.}\ \bibnamefont
  {Sandratskii}},\ }\href {\doibase 10.1103/PhysRevB.96.024450} {\bibfield
  {journal} {\bibinfo  {journal} {Phys. Rev. B}\ }\textbf {\bibinfo {volume}
  {96}},\ \bibinfo {pages} {024450} (\bibinfo {year} {2017})}\BibitemShut
  {NoStop}%
\bibitem [{\citenamefont {Le\ifmmode \check{z}\else
  \v{z}\fi{}ai\ifmmode~\acute{c}\else \'{c}\fi{}}\ \emph
  {et~al.}(2013)\citenamefont {Le\ifmmode \check{z}\else
  \v{z}\fi{}ai\ifmmode~\acute{c}\else \'{c}\fi{}}, \citenamefont {Mavropoulos},
  \citenamefont {Bihlmayer},\ and\ \citenamefont {Bl\"ugel}}]{Lezaic-PRB-2013}%
  \BibitemOpen
  \bibfield  {author} {\bibinfo {author} {\bibfnamefont {M.}~\bibnamefont
  {Le\ifmmode \check{z}\else \v{z}\fi{}ai\ifmmode~\acute{c}\else \'{c}\fi{}}},
  \bibinfo {author} {\bibfnamefont {P.}~\bibnamefont {Mavropoulos}}, \bibinfo
  {author} {\bibfnamefont {G.}~\bibnamefont {Bihlmayer}}, \ and\ \bibinfo
  {author} {\bibfnamefont {S.}~\bibnamefont {Bl\"ugel}},\ }\href {\doibase
  10.1103/PhysRevB.88.134403} {\bibfield  {journal} {\bibinfo  {journal} {Phys.
  Rev. B}\ }\textbf {\bibinfo {volume} {88}},\ \bibinfo {pages} {134403}
  (\bibinfo {year} {2013})}\BibitemShut {NoStop}%
\bibitem [{\citenamefont {Liechtenstein}\ \emph {et~al.}(1987)\citenamefont
  {Liechtenstein}, \citenamefont {Katsnelson}, \citenamefont {Antropov},\ and\
  \citenamefont {Gubanov}}]{Liechtenstein1987}%
  \BibitemOpen
  \bibfield  {author} {\bibinfo {author} {\bibfnamefont {A.}~\bibnamefont
  {Liechtenstein}}, \bibinfo {author} {\bibfnamefont {M.}~\bibnamefont
  {Katsnelson}}, \bibinfo {author} {\bibfnamefont {V.}~\bibnamefont
  {Antropov}}, \ and\ \bibinfo {author} {\bibfnamefont {V.}~\bibnamefont
  {Gubanov}},\ }\href {\doibase https://doi.org/10.1016/0304-8853(87)90721-9}
  {\bibfield  {journal} {\bibinfo  {journal} {Journal of Magnetism and Magnetic
  Materials}\ }\textbf {\bibinfo {volume} {67}},\ \bibinfo {pages} {65 }
  (\bibinfo {year} {1987})}\BibitemShut {NoStop}%
\bibitem [{\citenamefont {Jansen}(1999)}]{PhysRevB.59.4699}%
  \BibitemOpen
  \bibfield  {author} {\bibinfo {author} {\bibfnamefont {H.~J.~F.}\
  \bibnamefont {Jansen}},\ }\href {\doibase 10.1103/PhysRevB.59.4699}
  {\bibfield  {journal} {\bibinfo  {journal} {Phys. Rev. B}\ }\textbf {\bibinfo
  {volume} {59}},\ \bibinfo {pages} {4699} (\bibinfo {year}
  {1999})}\BibitemShut {NoStop}%
\bibitem [{\citenamefont {Heinze}\ \emph {et~al.}(2011)\citenamefont {Heinze},
  \citenamefont {von Bergmann}, \citenamefont {Menzel}, \citenamefont {Brede},
  \citenamefont {Kubetzka}, \citenamefont {Wiesendanger}, \citenamefont
  {Bihlmayer},\ and\ \citenamefont {Bl\"{u}gel}}]{Heinze2011}%
  \BibitemOpen
  \bibfield  {author} {\bibinfo {author} {\bibfnamefont {S.}~\bibnamefont
  {Heinze}}, \bibinfo {author} {\bibfnamefont {K.}~\bibnamefont {von
  Bergmann}}, \bibinfo {author} {\bibfnamefont {M.}~\bibnamefont {Menzel}},
  \bibinfo {author} {\bibfnamefont {J.}~\bibnamefont {Brede}}, \bibinfo
  {author} {\bibfnamefont {A.}~\bibnamefont {Kubetzka}}, \bibinfo {author}
  {\bibfnamefont {R.}~\bibnamefont {Wiesendanger}}, \bibinfo {author}
  {\bibfnamefont {G.}~\bibnamefont {Bihlmayer}}, \ and\ \bibinfo {author}
  {\bibfnamefont {S.}~\bibnamefont {Bl\"{u}gel}},\ }\href {\doibase
  10.1038/nphys2045} {\bibfield  {journal} {\bibinfo  {journal} {Nat. Phys.}\
  }\textbf {\bibinfo {volume} {7}},\ \bibinfo {pages} {713} (\bibinfo {year}
  {2011})}\BibitemShut {NoStop}%
\bibitem [{\citenamefont {Kr\"onlein}\ \emph {et~al.}(2018)\citenamefont
  {Kr\"onlein}, \citenamefont {Schmitt}, \citenamefont {Hoffmann},
  \citenamefont {Kemmer}, \citenamefont {Seubert}, \citenamefont {Vogt},
  \citenamefont {K\"uspert}, \citenamefont {B\"ohme}, \citenamefont {Alonazi},
  \citenamefont {K\"ugel}, \citenamefont {Albrithen}, \citenamefont {Bode},
  \citenamefont {Bihlmayer},\ and\ \citenamefont
  {Bl\"ugel}}]{PhysRevLett.120.207202}%
  \BibitemOpen
  \bibfield  {author} {\bibinfo {author} {\bibfnamefont {A.}~\bibnamefont
  {Kr\"onlein}}, \bibinfo {author} {\bibfnamefont {M.}~\bibnamefont {Schmitt}},
  \bibinfo {author} {\bibfnamefont {M.}~\bibnamefont {Hoffmann}}, \bibinfo
  {author} {\bibfnamefont {J.}~\bibnamefont {Kemmer}}, \bibinfo {author}
  {\bibfnamefont {N.}~\bibnamefont {Seubert}}, \bibinfo {author} {\bibfnamefont
  {M.}~\bibnamefont {Vogt}}, \bibinfo {author} {\bibfnamefont {J.}~\bibnamefont
  {K\"uspert}}, \bibinfo {author} {\bibfnamefont {M.}~\bibnamefont {B\"ohme}},
  \bibinfo {author} {\bibfnamefont {B.}~\bibnamefont {Alonazi}}, \bibinfo
  {author} {\bibfnamefont {J.}~\bibnamefont {K\"ugel}}, \bibinfo {author}
  {\bibfnamefont {H.~A.}\ \bibnamefont {Albrithen}}, \bibinfo {author}
  {\bibfnamefont {M.}~\bibnamefont {Bode}}, \bibinfo {author} {\bibfnamefont
  {G.}~\bibnamefont {Bihlmayer}}, \ and\ \bibinfo {author} {\bibfnamefont
  {S.}~\bibnamefont {Bl\"ugel}},\ }\href {\doibase
  10.1103/PhysRevLett.120.207202} {\bibfield  {journal} {\bibinfo  {journal}
  {Phys. Rev. Lett.}\ }\textbf {\bibinfo {volume} {120}},\ \bibinfo {pages}
  {207202} (\bibinfo {year} {2018})}\BibitemShut {NoStop}%
\bibitem [{\citenamefont {Szunyogh}\ \emph {et~al.}(1994)\citenamefont
  {Szunyogh}, \citenamefont {\'{U}jfalussy}, \citenamefont {Weinberger},\ and\
  \citenamefont {Koll\'{a}r}}]{Szunyogh1}%
  \BibitemOpen
  \bibfield  {author} {\bibinfo {author} {\bibfnamefont {L.}~\bibnamefont
  {Szunyogh}}, \bibinfo {author} {\bibfnamefont {B.}~\bibnamefont
  {\'{U}jfalussy}}, \bibinfo {author} {\bibfnamefont {P.}~\bibnamefont
  {Weinberger}}, \ and\ \bibinfo {author} {\bibfnamefont {J.}~\bibnamefont
  {Koll\'{a}r}},\ }\href@noop {} {\bibfield  {journal} {\bibinfo  {journal}
  {Phys. Rev. B}\ }\textbf {\bibinfo {volume} {49}},\ \bibinfo {pages} {2721}
  (\bibinfo {year} {1994})}\BibitemShut {NoStop}%
\bibitem [{\citenamefont {Zeller}\ \emph {et~al.}(1995)\citenamefont {Zeller},
  \citenamefont {Dederichs}, \citenamefont {\'{U}jfalussy}, \citenamefont
  {Szunyogh},\ and\ \citenamefont {Weinberger}}]{Zeller}%
  \BibitemOpen
  \bibfield  {author} {\bibinfo {author} {\bibfnamefont {R.}~\bibnamefont
  {Zeller}}, \bibinfo {author} {\bibfnamefont {P.~H.}\ \bibnamefont
  {Dederichs}}, \bibinfo {author} {\bibfnamefont {B.}~\bibnamefont
  {\'{U}jfalussy}}, \bibinfo {author} {\bibfnamefont {L.}~\bibnamefont
  {Szunyogh}}, \ and\ \bibinfo {author} {\bibfnamefont {P.}~\bibnamefont
  {Weinberger}},\ }\href {\doibase 10.1103/PhysRevB.52.8807} {\bibfield
  {journal} {\bibinfo  {journal} {Phys. Rev. B}\ }\textbf {\bibinfo {volume}
  {52}},\ \bibinfo {pages} {8807} (\bibinfo {year} {1995})}\BibitemShut
  {NoStop}%
\bibitem [{\citenamefont {Szunyogh}\ \emph
  {et~al.}(1995{\natexlab{a}})\citenamefont {Szunyogh}, \citenamefont
  {\'Ujfalussy},\ and\ \citenamefont {Weinberger}}]{Szunyogh2}%
  \BibitemOpen
  \bibfield  {author} {\bibinfo {author} {\bibfnamefont {L.}~\bibnamefont
  {Szunyogh}}, \bibinfo {author} {\bibfnamefont {B.}~\bibnamefont
  {\'Ujfalussy}}, \ and\ \bibinfo {author} {\bibfnamefont {P.}~\bibnamefont
  {Weinberger}},\ }\href {\doibase 10.1103/PhysRevB.51.9552} {\bibfield
  {journal} {\bibinfo  {journal} {Phys. Rev. B}\ }\textbf {\bibinfo {volume}
  {51}},\ \bibinfo {pages} {9552} (\bibinfo {year}
  {1995}{\natexlab{a}})}\BibitemShut {NoStop}%
\bibitem [{\citenamefont {Ferriani}\ \emph {et~al.}(2008)\citenamefont
  {Ferriani}, \citenamefont {von Bergmann}, \citenamefont {Vedmedenko},
  \citenamefont {Heinze}, \citenamefont {Bode}, \citenamefont {Heide},
  \citenamefont {Bihlmayer}, \citenamefont {Bl\"ugel},\ and\ \citenamefont
  {Wiesendanger}}]{PhysRevLett.101.027201}%
  \BibitemOpen
  \bibfield  {author} {\bibinfo {author} {\bibfnamefont {P.}~\bibnamefont
  {Ferriani}}, \bibinfo {author} {\bibfnamefont {K.}~\bibnamefont {von
  Bergmann}}, \bibinfo {author} {\bibfnamefont {E.~Y.}\ \bibnamefont
  {Vedmedenko}}, \bibinfo {author} {\bibfnamefont {S.}~\bibnamefont {Heinze}},
  \bibinfo {author} {\bibfnamefont {M.}~\bibnamefont {Bode}}, \bibinfo {author}
  {\bibfnamefont {M.}~\bibnamefont {Heide}}, \bibinfo {author} {\bibfnamefont
  {G.}~\bibnamefont {Bihlmayer}}, \bibinfo {author} {\bibfnamefont
  {S.}~\bibnamefont {Bl\"ugel}}, \ and\ \bibinfo {author} {\bibfnamefont
  {R.}~\bibnamefont {Wiesendanger}},\ }\href {\doibase
  10.1103/PhysRevLett.101.027201} {\bibfield  {journal} {\bibinfo  {journal}
  {Phys. Rev. Lett.}\ }\textbf {\bibinfo {volume} {101}},\ \bibinfo {pages}
  {027201} (\bibinfo {year} {2008})}\BibitemShut {NoStop}%
\bibitem [{\citenamefont {Dup\'{e}}\ \emph {et~al.}(2014)\citenamefont
  {Dup\'{e}}, \citenamefont {Hoffmann}, \citenamefont {Paillard},\ and\
  \citenamefont {Heinze}}]{Dupe2014}%
  \BibitemOpen
  \bibfield  {author} {\bibinfo {author} {\bibfnamefont {B.}~\bibnamefont
  {Dup\'{e}}}, \bibinfo {author} {\bibfnamefont {M.}~\bibnamefont {Hoffmann}},
  \bibinfo {author} {\bibfnamefont {C.}~\bibnamefont {Paillard}}, \ and\
  \bibinfo {author} {\bibfnamefont {S.}~\bibnamefont {Heinze}},\ }\href
  {http://dx.doi.org/10.1038/ncomms5030} {\bibfield  {journal} {\bibinfo
  {journal} {Nat. Commun.}\ }\textbf {\bibinfo {volume} {5}},\ \bibinfo {pages}
  {4030} (\bibinfo {year} {2014})}\BibitemShut {NoStop}%
\bibitem [{\citenamefont {Vida}\ \emph {et~al.}(2016)\citenamefont {Vida},
  \citenamefont {Simon}, \citenamefont {R\'{o}zsa}, \citenamefont
  {Palot\'{a}s},\ and\ \citenamefont {Szunyogh}}]{PhysRevB.94.214422}%
  \BibitemOpen
  \bibfield  {author} {\bibinfo {author} {\bibfnamefont {G.~J.}\ \bibnamefont
  {Vida}}, \bibinfo {author} {\bibfnamefont {E.}~\bibnamefont {Simon}},
  \bibinfo {author} {\bibfnamefont {L.}~\bibnamefont {R\'{o}zsa}}, \bibinfo
  {author} {\bibfnamefont {K.}~\bibnamefont {Palot\'{a}s}}, \ and\ \bibinfo
  {author} {\bibfnamefont {L.}~\bibnamefont {Szunyogh}},\ }\href {\doibase
  10.1103/PhysRevB.94.214422} {\bibfield  {journal} {\bibinfo  {journal} {Phys.
  Rev. B}\ }\textbf {\bibinfo {volume} {94}},\ \bibinfo {pages} {214422}
  (\bibinfo {year} {2016})}\BibitemShut {NoStop}%
\bibitem [{\citenamefont {Zimmermann}\ \emph {et~al.}(2019)\citenamefont
  {Zimmermann}, \citenamefont {Bihlmayer}, \citenamefont {B\"ottcher},
  \citenamefont {Bouhassoune}, \citenamefont {Lounis}, \citenamefont {Sinova},
  \citenamefont {Heinze}, \citenamefont {Bl\"ugel},\ and\ \citenamefont
  {Dup\'e}}]{PhysRevB.99.214426}%
  \BibitemOpen
  \bibfield  {author} {\bibinfo {author} {\bibfnamefont {B.}~\bibnamefont
  {Zimmermann}}, \bibinfo {author} {\bibfnamefont {G.}~\bibnamefont
  {Bihlmayer}}, \bibinfo {author} {\bibfnamefont {M.}~\bibnamefont
  {B\"ottcher}}, \bibinfo {author} {\bibfnamefont {M.}~\bibnamefont
  {Bouhassoune}}, \bibinfo {author} {\bibfnamefont {S.}~\bibnamefont {Lounis}},
  \bibinfo {author} {\bibfnamefont {J.}~\bibnamefont {Sinova}}, \bibinfo
  {author} {\bibfnamefont {S.}~\bibnamefont {Heinze}}, \bibinfo {author}
  {\bibfnamefont {S.}~\bibnamefont {Bl\"ugel}}, \ and\ \bibinfo {author}
  {\bibfnamefont {B.}~\bibnamefont {Dup\'e}},\ }\href {\doibase
  10.1103/PhysRevB.99.214426} {\bibfield  {journal} {\bibinfo  {journal} {Phys.
  Rev. B}\ }\textbf {\bibinfo {volume} {99}},\ \bibinfo {pages} {214426}
  (\bibinfo {year} {2019})}\BibitemShut {NoStop}%
\bibitem [{\citenamefont {Hrabec}\ \emph {et~al.}(2014)\citenamefont {Hrabec},
  \citenamefont {Porter}, \citenamefont {Wells}, \citenamefont {Benitez},
  \citenamefont {Burnell}, \citenamefont {McVitie}, \citenamefont {McGrouther},
  \citenamefont {Moore},\ and\ \citenamefont {Marrows}}]{Hrabec-PRB-2014}%
  \BibitemOpen
  \bibfield  {author} {\bibinfo {author} {\bibfnamefont {A.}~\bibnamefont
  {Hrabec}}, \bibinfo {author} {\bibfnamefont {N.~A.}\ \bibnamefont {Porter}},
  \bibinfo {author} {\bibfnamefont {A.}~\bibnamefont {Wells}}, \bibinfo
  {author} {\bibfnamefont {M.~J.}\ \bibnamefont {Benitez}}, \bibinfo {author}
  {\bibfnamefont {G.}~\bibnamefont {Burnell}}, \bibinfo {author} {\bibfnamefont
  {S.}~\bibnamefont {McVitie}}, \bibinfo {author} {\bibfnamefont
  {D.}~\bibnamefont {McGrouther}}, \bibinfo {author} {\bibfnamefont {T.~A.}\
  \bibnamefont {Moore}}, \ and\ \bibinfo {author} {\bibfnamefont {C.~H.}\
  \bibnamefont {Marrows}},\ }\href {\doibase 10.1103/PhysRevB.90.020402}
  {\bibfield  {journal} {\bibinfo  {journal} {Phys. Rev. B}\ }\textbf {\bibinfo
  {volume} {90}},\ \bibinfo {pages} {020402} (\bibinfo {year}
  {2014})}\BibitemShut {NoStop}%
\bibitem [{\citenamefont {Karayev}\ \emph {et~al.}(2019)\citenamefont
  {Karayev}, \citenamefont {Murray}, \citenamefont {Khadka}, \citenamefont
  {Thapaliya}, \citenamefont {Liu},\ and\ \citenamefont
  {Huang}}]{PhysRevMaterials.3.041401}%
  \BibitemOpen
  \bibfield  {author} {\bibinfo {author} {\bibfnamefont {S.}~\bibnamefont
  {Karayev}}, \bibinfo {author} {\bibfnamefont {P.~D.}\ \bibnamefont {Murray}},
  \bibinfo {author} {\bibfnamefont {D.}~\bibnamefont {Khadka}}, \bibinfo
  {author} {\bibfnamefont {T.~R.}\ \bibnamefont {Thapaliya}}, \bibinfo {author}
  {\bibfnamefont {K.}~\bibnamefont {Liu}}, \ and\ \bibinfo {author}
  {\bibfnamefont {S.~X.}\ \bibnamefont {Huang}},\ }\href {\doibase
  10.1103/PhysRevMaterials.3.041401} {\bibfield  {journal} {\bibinfo  {journal}
  {Phys. Rev. Materials}\ }\textbf {\bibinfo {volume} {3}},\ \bibinfo {pages}
  {041401} (\bibinfo {year} {2019})}\BibitemShut {NoStop}%
\bibitem [{\citenamefont {Han}\ \emph {et~al.}(2019)\citenamefont {Han},
  \citenamefont {Lee}, \citenamefont {Hanke}, \citenamefont {Mokrousov},
  \citenamefont {Kim}, \citenamefont {Yoo}, \citenamefont {van Hees},
  \citenamefont {Kim}, \citenamefont {Lavrijsen}, \citenamefont {You},
  \citenamefont {Swagten}, \citenamefont {Jung},\ and\ \citenamefont
  {Kl\"aui}}]{Han2019}%
  \BibitemOpen
  \bibfield  {author} {\bibinfo {author} {\bibfnamefont {D.-S.}\ \bibnamefont
  {Han}}, \bibinfo {author} {\bibfnamefont {K.}~\bibnamefont {Lee}}, \bibinfo
  {author} {\bibfnamefont {J.-P.}\ \bibnamefont {Hanke}}, \bibinfo {author}
  {\bibfnamefont {Y.}~\bibnamefont {Mokrousov}}, \bibinfo {author}
  {\bibfnamefont {K.-W.}\ \bibnamefont {Kim}}, \bibinfo {author} {\bibfnamefont
  {W.}~\bibnamefont {Yoo}}, \bibinfo {author} {\bibfnamefont {Y.~L.~W.}\
  \bibnamefont {van Hees}}, \bibinfo {author} {\bibfnamefont {T.-W.}\
  \bibnamefont {Kim}}, \bibinfo {author} {\bibfnamefont {R.}~\bibnamefont
  {Lavrijsen}}, \bibinfo {author} {\bibfnamefont {C.-Y.}\ \bibnamefont {You}},
  \bibinfo {author} {\bibfnamefont {H.~J.~M.}\ \bibnamefont {Swagten}},
  \bibinfo {author} {\bibfnamefont {M.-H.}\ \bibnamefont {Jung}}, \ and\
  \bibinfo {author} {\bibfnamefont {M.}~\bibnamefont {Kl\"aui}},\ }\href
  {\doibase 10.1038/s41563-019-0370-z} {\bibfield  {journal} {\bibinfo
  {journal} {Nat. Mat.}\ }\textbf {\bibinfo {volume} {18}},\ \bibinfo {pages}
  {703} (\bibinfo {year} {2019})}\BibitemShut {NoStop}%
\bibitem [{\citenamefont {Soumyanarayanan}\ \emph {et~al.}(2017)\citenamefont
  {Soumyanarayanan}, \citenamefont {Raju}, \citenamefont {Gonzalez~Oyarce},
  \citenamefont {Tan}, \citenamefont {Im}, \citenamefont {Petrovi\'{c}},
  \citenamefont {Ho}, \citenamefont {Khoo}, \citenamefont {Tran}, \citenamefont
  {Gan}, \citenamefont {Ernult},\ and\ \citenamefont
  {Panagopoulos}}]{Soumyanarayanan2017}%
  \BibitemOpen
  \bibfield  {author} {\bibinfo {author} {\bibfnamefont {A.}~\bibnamefont
  {Soumyanarayanan}}, \bibinfo {author} {\bibfnamefont {M.}~\bibnamefont
  {Raju}}, \bibinfo {author} {\bibfnamefont {A.~L.}\ \bibnamefont
  {Gonzalez~Oyarce}}, \bibinfo {author} {\bibfnamefont {A.~K.~C.}\ \bibnamefont
  {Tan}}, \bibinfo {author} {\bibfnamefont {M.-Y.}\ \bibnamefont {Im}},
  \bibinfo {author} {\bibfnamefont {A.~P.}\ \bibnamefont {Petrovi\'{c}}},
  \bibinfo {author} {\bibfnamefont {P.}~\bibnamefont {Ho}}, \bibinfo {author}
  {\bibfnamefont {K.~H.}\ \bibnamefont {Khoo}}, \bibinfo {author}
  {\bibfnamefont {M.}~\bibnamefont {Tran}}, \bibinfo {author} {\bibfnamefont
  {C.~K.}\ \bibnamefont {Gan}}, \bibinfo {author} {\bibfnamefont
  {F.}~\bibnamefont {Ernult}}, \ and\ \bibinfo {author} {\bibfnamefont
  {C.}~\bibnamefont {Panagopoulos}},\ }\href {https://doi.org/10.1038/nmat4934}
  {\bibfield  {journal} {\bibinfo  {journal} {Nat. Mater.}\ }\textbf {\bibinfo
  {volume} {16}},\ \bibinfo {pages} {898} (\bibinfo {year} {2017})}\BibitemShut
  {NoStop}%
\bibitem [{\citenamefont {Yagil}\ \emph {et~al.}(2018)\citenamefont {Yagil},
  \citenamefont {Almoalem}, \citenamefont {Soumyanarayanan}, \citenamefont
  {Tan}, \citenamefont {Raju}, \citenamefont {Panagopoulos},\ and\
  \citenamefont {Auslaender}}]{Yagil_APL}%
  \BibitemOpen
  \bibfield  {author} {\bibinfo {author} {\bibfnamefont {A.}~\bibnamefont
  {Yagil}}, \bibinfo {author} {\bibfnamefont {A.}~\bibnamefont {Almoalem}},
  \bibinfo {author} {\bibfnamefont {A.}~\bibnamefont {Soumyanarayanan}},
  \bibinfo {author} {\bibfnamefont {A.~K.~C.}\ \bibnamefont {Tan}}, \bibinfo
  {author} {\bibfnamefont {M.}~\bibnamefont {Raju}}, \bibinfo {author}
  {\bibfnamefont {C.}~\bibnamefont {Panagopoulos}}, \ and\ \bibinfo {author}
  {\bibfnamefont {O.~M.}\ \bibnamefont {Auslaender}},\ }\href {\doibase
  10.1063/1.5027602} {\bibfield  {journal} {\bibinfo  {journal} {Appl. Phys.
  Lett.}\ }\textbf {\bibinfo {volume} {112}},\ \bibinfo {pages} {192403}
  (\bibinfo {year} {2018})}\BibitemShut {NoStop}%
\bibitem [{\citenamefont {Raju}\ \emph {et~al.}(2019)\citenamefont {Raju},
  \citenamefont {Yagil}, \citenamefont {Soumyanarayanan}, \citenamefont {Tan},
  \citenamefont {Almoalem}, \citenamefont {Ma}, \citenamefont {Auslaender},\
  and\ \citenamefont {Panagopoulos}}]{Raju2019}%
  \BibitemOpen
  \bibfield  {author} {\bibinfo {author} {\bibfnamefont {M.}~\bibnamefont
  {Raju}}, \bibinfo {author} {\bibfnamefont {A.}~\bibnamefont {Yagil}},
  \bibinfo {author} {\bibfnamefont {A.}~\bibnamefont {Soumyanarayanan}},
  \bibinfo {author} {\bibfnamefont {A.~K.~C.}\ \bibnamefont {Tan}}, \bibinfo
  {author} {\bibfnamefont {A.}~\bibnamefont {Almoalem}}, \bibinfo {author}
  {\bibfnamefont {F.}~\bibnamefont {Ma}}, \bibinfo {author} {\bibfnamefont
  {O.~M.}\ \bibnamefont {Auslaender}}, \ and\ \bibinfo {author} {\bibfnamefont
  {C.}~\bibnamefont {Panagopoulos}},\ }\href {\doibase
  10.1038/s41467-018-08041-9} {\bibfield  {journal} {\bibinfo  {journal} {Nat.
  Commun.}\ }\textbf {\bibinfo {volume} {10}},\ \bibinfo {pages} {696}
  (\bibinfo {year} {2019})}\BibitemShut {NoStop}%
\bibitem [{\citenamefont {Dup\'{e}}\ \emph {et~al.}(2016)\citenamefont
  {Dup\'{e}}, \citenamefont {Bihlmayer}, \citenamefont {B{\"o}ttcher},
  \citenamefont {Bl\"ugel},\ and\ \citenamefont {Heinze}}]{Dupe2016}%
  \BibitemOpen
  \bibfield  {author} {\bibinfo {author} {\bibfnamefont {B.}~\bibnamefont
  {Dup\'{e}}}, \bibinfo {author} {\bibfnamefont {G.}~\bibnamefont {Bihlmayer}},
  \bibinfo {author} {\bibfnamefont {M.}~\bibnamefont {B{\"o}ttcher}}, \bibinfo
  {author} {\bibfnamefont {S.}~\bibnamefont {Bl\"ugel}}, \ and\ \bibinfo
  {author} {\bibfnamefont {S.}~\bibnamefont {Heinze}},\ }\href
  {https://doi.org/10.1038/ncomms11779} {\bibfield  {journal} {\bibinfo
  {journal} {Nat. Commun.}\ }\textbf {\bibinfo {volume} {7}},\ \bibinfo {pages}
  {11779 EP } (\bibinfo {year} {2016})}\BibitemShut {NoStop}%
\bibitem [{\citenamefont {Ewald}(1921)}]{Ewald-1921}%
  \BibitemOpen
  \bibfield  {author} {\bibinfo {author} {\bibfnamefont {P.~P.}\ \bibnamefont
  {Ewald}},\ }\href {\doibase 10.1002/andp.19213690304} {\bibfield  {journal}
  {\bibinfo  {journal} {Annalen der Physik}\ }\textbf {\bibinfo {volume}
  {369}},\ \bibinfo {pages} {253} (\bibinfo {year} {1921})}\BibitemShut
  {NoStop}%
\bibitem [{\citenamefont {Ham}\ and\ \citenamefont
  {Segall}(1961)}]{Ham-Segall-1961}%
  \BibitemOpen
  \bibfield  {author} {\bibinfo {author} {\bibfnamefont {F.~S.}\ \bibnamefont
  {Ham}}\ and\ \bibinfo {author} {\bibfnamefont {B.}~\bibnamefont {Segall}},\
  }\href {\doibase 10.1103/PhysRev.124.1786} {\bibfield  {journal} {\bibinfo
  {journal} {Phys. Rev.}\ }\textbf {\bibinfo {volume} {124}},\ \bibinfo {pages}
  {1786} (\bibinfo {year} {1961})}\BibitemShut {NoStop}%
\bibitem [{\citenamefont {Kambe}(1967{\natexlab{a}})}]{Kambe1}%
  \BibitemOpen
  \bibfield  {author} {\bibinfo {author} {\bibfnamefont {K.}~\bibnamefont
  {Kambe}},\ }\href@noop {} {\bibfield  {journal} {\bibinfo  {journal} {Z.
  Naturforschung}\ }\textbf {\bibinfo {volume} {22a}},\ \bibinfo {pages} {322}
  (\bibinfo {year} {1967}{\natexlab{a}})}\BibitemShut {NoStop}%
\bibitem [{\citenamefont {Kambe}(1967{\natexlab{b}})}]{Kambe2}%
  \BibitemOpen
  \bibfield  {author} {\bibinfo {author} {\bibfnamefont {K.}~\bibnamefont
  {Kambe}},\ }\href@noop {} {\bibfield  {journal} {\bibinfo  {journal} {Z.
  Naturforschung}\ }\textbf {\bibinfo {volume} {22a}},\ \bibinfo {pages} {422}
  (\bibinfo {year} {1967}{\natexlab{b}})}\BibitemShut {NoStop}%
\bibitem [{\citenamefont {Kambe}(1968)}]{Kambe3}%
  \BibitemOpen
  \bibfield  {author} {\bibinfo {author} {\bibfnamefont {K.}~\bibnamefont
  {Kambe}},\ }\href@noop {} {\bibfield  {journal} {\bibinfo  {journal} {Z.
  Naturforschung}\ }\textbf {\bibinfo {volume} {23a}},\ \bibinfo {pages} {1280}
  (\bibinfo {year} {1968})}\BibitemShut {NoStop}%
\bibitem [{\citenamefont {Hasselberg}\ \emph {et~al.}(2015)\citenamefont
  {Hasselberg}, \citenamefont {Yanes}, \citenamefont {Hinzke}, \citenamefont
  {Sessi}, \citenamefont {Bode}, \citenamefont {Szunyogh},\ and\ \citenamefont
  {Nowak}}]{Rocio-2015}%
  \BibitemOpen
  \bibfield  {author} {\bibinfo {author} {\bibfnamefont {G.}~\bibnamefont
  {Hasselberg}}, \bibinfo {author} {\bibfnamefont {R.}~\bibnamefont {Yanes}},
  \bibinfo {author} {\bibfnamefont {D.}~\bibnamefont {Hinzke}}, \bibinfo
  {author} {\bibfnamefont {P.}~\bibnamefont {Sessi}}, \bibinfo {author}
  {\bibfnamefont {M.}~\bibnamefont {Bode}}, \bibinfo {author} {\bibfnamefont
  {L.}~\bibnamefont {Szunyogh}}, \ and\ \bibinfo {author} {\bibfnamefont
  {U.}~\bibnamefont {Nowak}},\ }\href {\doibase 10.1103/PhysRevB.91.064402}
  {\bibfield  {journal} {\bibinfo  {journal} {Phys. Rev. B}\ }\textbf {\bibinfo
  {volume} {91}},\ \bibinfo {pages} {064402} (\bibinfo {year}
  {2015})}\BibitemShut {NoStop}%
\bibitem [{\citenamefont {Lloyd}(1967)}]{Lloyd_1967}%
  \BibitemOpen
  \bibfield  {author} {\bibinfo {author} {\bibfnamefont {P.}~\bibnamefont
  {Lloyd}},\ }\href {\doibase 10.1088/0370-1328/90/1/324} {\bibfield  {journal}
  {\bibinfo  {journal} {Proc. Phys. Soc.}\ }\textbf {\bibinfo {volume} {90}},\
  \bibinfo {pages} {217} (\bibinfo {year} {1967})}\BibitemShut {NoStop}%
\bibitem [{\citenamefont {Simon}\ \emph {et~al.}(2018)\citenamefont {Simon},
  \citenamefont {R\'ozsa}, \citenamefont {Palot\'as},\ and\ \citenamefont
  {Szunyogh}}]{PhysRevB.97.134405}%
  \BibitemOpen
  \bibfield  {author} {\bibinfo {author} {\bibfnamefont {E.}~\bibnamefont
  {Simon}}, \bibinfo {author} {\bibfnamefont {L.}~\bibnamefont {R\'ozsa}},
  \bibinfo {author} {\bibfnamefont {K.}~\bibnamefont {Palot\'as}}, \ and\
  \bibinfo {author} {\bibfnamefont {L.}~\bibnamefont {Szunyogh}},\ }\href
  {\doibase 10.1103/PhysRevB.97.134405} {\bibfield  {journal} {\bibinfo
  {journal} {Phys. Rev. B}\ }\textbf {\bibinfo {volume} {97}},\ \bibinfo
  {pages} {134405} (\bibinfo {year} {2018})}\BibitemShut {NoStop}%
\bibitem [{\citenamefont {Schweflinghaus}\ \emph {et~al.}(2016)\citenamefont
  {Schweflinghaus}, \citenamefont {Zimmermann}, \citenamefont {Heide},
  \citenamefont {Bihlmayer},\ and\ \citenamefont
  {Bl\"ugel}}]{Schweflinghaus-PRB2016}%
  \BibitemOpen
  \bibfield  {author} {\bibinfo {author} {\bibfnamefont {B.}~\bibnamefont
  {Schweflinghaus}}, \bibinfo {author} {\bibfnamefont {B.}~\bibnamefont
  {Zimmermann}}, \bibinfo {author} {\bibfnamefont {M.}~\bibnamefont {Heide}},
  \bibinfo {author} {\bibfnamefont {G.}~\bibnamefont {Bihlmayer}}, \ and\
  \bibinfo {author} {\bibfnamefont {S.}~\bibnamefont {Bl\"ugel}},\ }\href
  {\doibase 10.1103/PhysRevB.94.024403} {\bibfield  {journal} {\bibinfo
  {journal} {Phys. Rev. B}\ }\textbf {\bibinfo {volume} {94}},\ \bibinfo
  {pages} {024403} (\bibinfo {year} {2016})}\BibitemShut {NoStop}%
\bibitem [{\citenamefont {Freimuth}\ \emph {et~al.}(2014)\citenamefont
  {Freimuth}, \citenamefont {Bl\"{u}gel},\ and\ \citenamefont
  {Mokrousov}}]{Freimuth2014}%
  \BibitemOpen
  \bibfield  {author} {\bibinfo {author} {\bibfnamefont {F.}~\bibnamefont
  {Freimuth}}, \bibinfo {author} {\bibfnamefont {S.}~\bibnamefont
  {Bl\"{u}gel}}, \ and\ \bibinfo {author} {\bibfnamefont {Y.}~\bibnamefont
  {Mokrousov}},\ }\href {http://stacks.iop.org/0953-8984/26/i=10/a=104202}
  {\bibfield  {journal} {\bibinfo  {journal} {J. Phys.: Condens. Matter}\
  }\textbf {\bibinfo {volume} {26}},\ \bibinfo {pages} {104202} (\bibinfo
  {year} {2014})}\BibitemShut {NoStop}%
\bibitem [{\citenamefont {Daalderop}\ \emph {et~al.}(1990)\citenamefont
  {Daalderop}, \citenamefont {Kelly},\ and\ \citenamefont
  {Schuurmans}}]{PhysRevB.41.11919}%
  \BibitemOpen
  \bibfield  {author} {\bibinfo {author} {\bibfnamefont {G.~H.~O.}\
  \bibnamefont {Daalderop}}, \bibinfo {author} {\bibfnamefont {P.~J.}\
  \bibnamefont {Kelly}}, \ and\ \bibinfo {author} {\bibfnamefont {M.~F.~H.}\
  \bibnamefont {Schuurmans}},\ }\href {\doibase 10.1103/PhysRevB.41.11919}
  {\bibfield  {journal} {\bibinfo  {journal} {Phys. Rev. B}\ }\textbf {\bibinfo
  {volume} {41}},\ \bibinfo {pages} {11919} (\bibinfo {year}
  {1990})}\BibitemShut {NoStop}%
\bibitem [{\citenamefont {Szunyogh}\ \emph
  {et~al.}(1995{\natexlab{b}})\citenamefont {Szunyogh}, \citenamefont
  {\'Ujfalussy},\ and\ \citenamefont {Weinberger}}]{PhysRevB.51.9552}%
  \BibitemOpen
  \bibfield  {author} {\bibinfo {author} {\bibfnamefont {L.}~\bibnamefont
  {Szunyogh}}, \bibinfo {author} {\bibfnamefont {B.}~\bibnamefont
  {\'Ujfalussy}}, \ and\ \bibinfo {author} {\bibfnamefont {P.}~\bibnamefont
  {Weinberger}},\ }\href {\doibase 10.1103/PhysRevB.51.9552} {\bibfield
  {journal} {\bibinfo  {journal} {Phys. Rev. B}\ }\textbf {\bibinfo {volume}
  {51}},\ \bibinfo {pages} {9552} (\bibinfo {year}
  {1995}{\natexlab{b}})}\BibitemShut {NoStop}%
\bibitem [{\citenamefont {Vosko}\ \emph {et~al.}(1980)\citenamefont {Vosko},
  \citenamefont {Wilk},\ and\ \citenamefont {Nusair}}]{vosko}%
  \BibitemOpen
  \bibfield  {author} {\bibinfo {author} {\bibfnamefont {S.~H.}\ \bibnamefont
  {Vosko}}, \bibinfo {author} {\bibfnamefont {L.}~\bibnamefont {Wilk}}, \ and\
  \bibinfo {author} {\bibfnamefont {M.}~\bibnamefont {Nusair}},\ }\href@noop {}
  {\bibfield  {journal} {\bibinfo  {journal} {Can. J. Phys.}\ }\textbf
  {\bibinfo {volume} {58}},\ \bibinfo {pages} {1200} (\bibinfo {year}
  {1980})}\BibitemShut {NoStop}%
\bibitem [{\citenamefont {Kresse}\ and\ \citenamefont
  {Furthm\"{u}ller}(1996{\natexlab{a}})}]{Kresse199615}%
  \BibitemOpen
  \bibfield  {author} {\bibinfo {author} {\bibfnamefont {G.}~\bibnamefont
  {Kresse}}\ and\ \bibinfo {author} {\bibfnamefont {J.}~\bibnamefont
  {Furthm\"{u}ller}},\ }\href@noop {} {\bibfield  {journal} {\bibinfo
  {journal} {Comput. Mater. Sci.}\ }\textbf {\bibinfo {volume} {6}},\ \bibinfo
  {pages} {15 } (\bibinfo {year} {1996}{\natexlab{a}})}\BibitemShut {NoStop}%
\bibitem [{\citenamefont {Kresse}\ and\ \citenamefont
  {Furthm\"{u}ller}(1996{\natexlab{b}})}]{Kresse-PRB}%
  \BibitemOpen
  \bibfield  {author} {\bibinfo {author} {\bibfnamefont {G.}~\bibnamefont
  {Kresse}}\ and\ \bibinfo {author} {\bibfnamefont {J.}~\bibnamefont
  {Furthm\"{u}ller}},\ }\href@noop {} {\bibfield  {journal} {\bibinfo
  {journal} {Phys. Rev. B}\ }\textbf {\bibinfo {volume} {54}},\ \bibinfo
  {pages} {11169} (\bibinfo {year} {1996}{\natexlab{b}})}\BibitemShut {NoStop}%
\bibitem [{\citenamefont {Hafner}(2008)}]{Hafner}%
  \BibitemOpen
  \bibfield  {author} {\bibinfo {author} {\bibfnamefont {J.}~\bibnamefont
  {Hafner}},\ }\href@noop {} {\bibfield  {journal} {\bibinfo  {journal} {J.
  Comput. Chem.}\ }\textbf {\bibinfo {volume} {29}},\ \bibinfo {pages} {2044}
  (\bibinfo {year} {2008})}\BibitemShut {NoStop}%
\bibitem [{\citenamefont {Nandy}\ \emph {et~al.}(2016)\citenamefont {Nandy},
  \citenamefont {Kiselev},\ and\ \citenamefont
  {Bl\"ugel}}]{PhysRevLett.116.177202}%
  \BibitemOpen
  \bibfield  {author} {\bibinfo {author} {\bibfnamefont {A.~K.}\ \bibnamefont
  {Nandy}}, \bibinfo {author} {\bibfnamefont {N.~S.}\ \bibnamefont {Kiselev}},
  \ and\ \bibinfo {author} {\bibfnamefont {S.}~\bibnamefont {Bl\"ugel}},\
  }\href {\doibase 10.1103/PhysRevLett.116.177202} {\bibfield  {journal}
  {\bibinfo  {journal} {Phys. Rev. Lett.}\ }\textbf {\bibinfo {volume} {116}},\
  \bibinfo {pages} {177202} (\bibinfo {year} {2016})}\BibitemShut {NoStop}%
\bibitem [{\citenamefont {Ferriani}\ \emph {et~al.}(2005)\citenamefont
  {Ferriani}, \citenamefont {Heinze}, \citenamefont {Bihlmayer},\ and\
  \citenamefont {Bl\"ugel}}]{PhysRevB.72.024452}%
  \BibitemOpen
  \bibfield  {author} {\bibinfo {author} {\bibfnamefont {P.}~\bibnamefont
  {Ferriani}}, \bibinfo {author} {\bibfnamefont {S.}~\bibnamefont {Heinze}},
  \bibinfo {author} {\bibfnamefont {G.}~\bibnamefont {Bihlmayer}}, \ and\
  \bibinfo {author} {\bibfnamefont {S.}~\bibnamefont {Bl\"ugel}},\ }\href
  {\doibase 10.1103/PhysRevB.72.024452} {\bibfield  {journal} {\bibinfo
  {journal} {Phys. Rev. B}\ }\textbf {\bibinfo {volume} {72}},\ \bibinfo
  {pages} {024452} (\bibinfo {year} {2005})}\BibitemShut {NoStop}%
\bibitem [{\citenamefont {den Broeder}\ \emph {et~al.}(1991)\citenamefont {den
  Broeder}, \citenamefont {Hoving},\ and\ \citenamefont
  {Bloemen}}]{DENBROEDER1991562}%
  \BibitemOpen
  \bibfield  {author} {\bibinfo {author} {\bibfnamefont {F.}~\bibnamefont {den
  Broeder}}, \bibinfo {author} {\bibfnamefont {W.}~\bibnamefont {Hoving}}, \
  and\ \bibinfo {author} {\bibfnamefont {P.}~\bibnamefont {Bloemen}},\ }\href
  {\doibase https://doi.org/10.1016/0304-8853(91)90404-X} {\bibfield  {journal}
  {\bibinfo  {journal} {J. Magn. Magn. Mater}\ }\textbf {\bibinfo {volume}
  {93}},\ \bibinfo {pages} {562 } (\bibinfo {year} {1991})}\BibitemShut
  {NoStop}%
\bibitem [{\citenamefont {Wu}\ \emph {et~al.}(2012)\citenamefont {Wu},
  \citenamefont {Khoo}, \citenamefont {Jhon}, \citenamefont {Meng},
  \citenamefont {Lua}, \citenamefont {Sbiaa},\ and\ \citenamefont
  {Gan}}]{Wu_2012}%
  \BibitemOpen
  \bibfield  {author} {\bibinfo {author} {\bibfnamefont {G.}~\bibnamefont
  {Wu}}, \bibinfo {author} {\bibfnamefont {K.~H.}\ \bibnamefont {Khoo}},
  \bibinfo {author} {\bibfnamefont {M.~H.}\ \bibnamefont {Jhon}}, \bibinfo
  {author} {\bibfnamefont {H.}~\bibnamefont {Meng}}, \bibinfo {author}
  {\bibfnamefont {S.~Y.~H.}\ \bibnamefont {Lua}}, \bibinfo {author}
  {\bibfnamefont {R.}~\bibnamefont {Sbiaa}}, \ and\ \bibinfo {author}
  {\bibfnamefont {C.~K.}\ \bibnamefont {Gan}},\ }\href {\doibase
  10.1209/0295-5075/99/17001} {\bibfield  {journal} {\bibinfo  {journal}
  {Europhys. Lett.}\ }\textbf {\bibinfo {volume} {99}},\ \bibinfo {pages}
  {17001} (\bibinfo {year} {2012})}\BibitemShut {NoStop}%
\end{thebibliography}%
%
% end file
\end{document}